\documentclass[prd, reprint, amsfonts, amssymb, amsmath, preprintnumbers, showpacs, nofootinbib, superscriptaddress]{revtex4-1}
\usepackage{graphicx, amsthm, multirow, subfigure}
\usepackage[citecolor=blue]{hyperref}
\usepackage[all]{hypcap}
\usepackage{url}
\newcommand{\equaref}[1]{Eq.~\ref{#1}}
\newcommand{\figref}[1]{Fig.~\ref{#1}}

\newcommand{\secref}[1]{Section~\ref{#1}}
\newcommand{\refref}[1]{Ref.~\cite{#1}}

\begin{document}

\title{Understanding the performance of the low energy neutrino factory: the dependence on baseline distance and stored-muon energy}

\author{Peter~Ballett}
\email{peter.ballett@durham.ac.uk}
\author{Silvia~Pascoli} 
\email{silvia.pascoli@durham.ac.uk}
\affiliation{Institute for Particle Physics Phenomenology, Department of Physics, Durham University, Durham, DH1 3LE, United Kingdom.}

\date{\today}

\begin{abstract}
 
We study the physics reach of a Low Energy Neutrino Factory
(LENF) and its dependence on the chosen baseline distance, $L$, and stored-muon
energy, $E_\mu$, in order to ascertain the configuration of the optimal LENF.
In particular, we study the performance of the LENF over a range of baseline
distances from $1000$~km to $4000$~km and stored-muon energies from $4$~GeV to
$25$~GeV, connecting the early studies of the LENF ($1300$~km, $4.5$~GeV) to
those of the conventional, high-energy neutrino factory design ($4000$~km and
$7000$~km, $25$~GeV).  
Three different magnetized detector options are considered: a Totally-Active
Scintillator Detector (TASD) and two models of a liquid-argon detector
distinguished by optimistic and conservative performance estimates. 
In order to compare the sensitivity of each set-up, we compute the full
$\delta$-dependent discovery contours for the determination of
$\theta_{13}\neq0$, $\delta_{CP}\notin\{0,\pi\}$ and $\text{sign}(\Delta m^2_{13})$.
For large values of $\theta_{13}$, 
as recently confirmed by the Daya Bay and RENO experiments, 
the LENF provides a strong discovery potential over
the majority of the $L$--$E_\mu$ parameter space and is a promising candidate for the
future generation of long baseline experiments aimed at discovering
CP-violation and the mass hierarchy, and at making a precise determination of
the oscillation parameters. 

\end{abstract}

\preprint{IPPP/11/69, DCPT/11/138, EURONU-WP6-12-46}

\maketitle

\section{Introduction}

Over the last 50 years, neutrino oscillations have advanced from a tentative
hypothesis to a well confirmed physical phenomenon and have been observed in
various channels and  over a wide range of energy scales. 
Until recently, the results of several experiments (see,
e.g.~\cite{Adamson:2011ig, *Abe:2010hy, *Wendell:2010md, *Aharmim:2011vm,
*Derbin:2010zz, *Abe:2008ee}) had only been able to provide us with
measurements of four of the six oscillation parameters: $\theta_{12}$,
$\theta_{23}$, $\Delta m^2_{21}$ and $|\Delta
m^2_{31}|$~\cite{GonzalezGarcia:2010er,Fogli:2011qn,Schwetz:2011qt,Schwetz:2011zk}.
This situation has changed over the last few months, with new experimental
searches \cite{Abe:2011sj,Adamson:2011qu,DeKerret:2011aa,An:2012eh,Ahn:2012nd}
succeeding in measuring the last mixing angle $\theta_{13}$ and confirming that
it is large with central values reported by Daya Bay and RENO of
$\sin^22\theta_{13} = 0.092$ and $\sin^22\theta_{13} = 0.113$,
respectively~\cite{An:2012eh,Ahn:2012nd}. 
The discovery of non-zero $\theta_{13}$ has increased the likelihood that the
resolution of the final uncertainties in the neutrino mixing parameters will be
possible within a medium-term experimental program. There are two particularly
important unknowns that will dominate this project: measuring the complex phase
$\delta$, which governs CP-violating effects in the leptonic sector and is
currently unconstrained, and determining the sign of $\Delta m^2_{31}$, which
dictates the type of neutrino mass-ordering and controls the enhancement of the
oscillation probabilities in the Earth due to matter effects for long-baseline
and atmospheric neutrinos. 
 
The values of the unknown neutrino mixing parameters may well turn out to
have decisive implications for potential extensions of the Standard Model and
for understanding the problem of flavor.  Improving our knowledge of these
parameters is one of the main aims of the next generation of long-baseline
oscillation searches.  These experiments are primarily aimed at studying the
sub-dominant oscillation probability $P(\nu_e \rightarrow \nu_\mu)$, or
$P(\nu_\mu \rightarrow \nu_e)$.  These probabilities have a complex dependence
on the oscillation parameters and their extraction from data suffers from the
problem of degeneracies
\cite{Fogli:1996pv,BurguetCastell:2001ez,Minakata:2001qm,Barger:2001yr,Huber:2002mx,Minakata:2002qi,Donini:2003vz,Aoki:2003kc,Yasuda:2004gu}.
Various strategies have been devised to weaken or resolve this
problem~\cite{Kajita:2001sb,BurguetCastell:2002qx,Minakata:2003ca}, for
example the energy dependence of the degenerate solutions can be exploited
using a wide-band beam~\cite{Diwan:2003bp,Huber:2010dx}, matter effects can be
used to lift the hierarchy degeneracy~\cite{Barger:1980tf,Freund:1999gy} whilst,
for the measurement of $\theta_{23}$ and its deviations from maximality, the
low-energy part of the spectrum is particularly important~\cite{Geer:2007kn} as
well as a precise determination of the disappearance probability.

Amongst future long-baseline neutrino experiments, the Neutrino Factory
(NF)~\cite{Geer:1997iz,DeRujula:1998hd,Bandyopadhyay:2007kx} is a leading
design with excellent physics reach.  At a NF, muons (or anti-muons) produced
via pion decay are accelerated to a common energy $E_\mu$
(originally around $25$~GeV)
and are injected into a storage ring. A significant number of high-gamma muons
will decay producing a highly collimated beam of neutrinos, whose spectral
content is known with high precision.  The decay of $\mu^-$ ($\mu^+$) in the
storage ring produces an initial beam with two neutrino components: $\nu_\mu$
and $ \overline{\nu}_e$ ($\overline{\nu}_\mu$ and $\nu_e$). At the detector,
two different signals will be present:
for instance with $\mu^-$ decays,
the {\it right-sign} muon events which derive from the observation of $\nu_\mu$
coming from the disappearance channel and the {\it wrong-sign} muon events
which are due to $\bar{\nu}_e \rightarrow \bar{\nu}_\mu$ oscillations.  It is
necessary to distinguish these two signals in order to reconstruct the
appearance probability. This requires a magnetized detector which is able to
distinguish $\mu^+$- and $\mu^-$-events.  The baseline choice for the
conventional NF~\cite{Choubey:2011zz} is a Magnetized Iron Neutrino Detector
(MIND) which provides excellent background rejection and very good energy
resolution but low detection efficiency for neutrinos with energies in the few
GeV range~\cite{Cervera:2010rz,Laing:2010zz}. Other options include a
magnetized Totally-Active Scintillator Detector
(TASD)~\cite{Bandyopadhyay:2007kx} or a detector based on the liquid-argon time
projection
chamber~\cite{Rubbia:1977zz,Cennini:1994br,Rubbia:2004tz,Rubbia:2009md}. The
latter detectors provide excellent efficiency for neutrinos with energies as
low as a few hundred MeV to $1$ GeV, excellent energy resolution and low
backgrounds but their magnetization is extremely
challenging~\cite{Bross:2007ts}.

Before the measurement of $\theta_{13}$, the baseline configuration of the NF~\cite{Choubey:2011zz} used muons with an
energy of 25~GeV and two different baselines, at approximately $4000$ and
$7500$~km, with two MIND detectors, a 100 kton one at the shorter baseline and
a 50 kton one at the `magic' baseline~\cite{Huber:2003ak, Smirnov:2006sm}. The
combination of two distinct baseline distances was designed to help resolve the
problem of parameter degeneracies by providing complementary information of the
oscillation probability at different points in parameter space. This set-up was
optimized for small values of $\theta_{13}$ and has been shown to have an
excellent physics reach to $\theta_{13}$, CP-violation and the mass
hierarchy~\cite{Choubey:2011zz, Huber:2006wb, Agarwalla:2010hk}.

A set-up, named the Low-Energy Neutrino Factory (LENF), has been proposed
as a more conservative option for large values of
$\theta_{13}$~\cite{Geer:2007kn,Bross:2007ts,FernandezMartinez:2010zza}.
The first proposal for the LENF used a single baseline of $1300$~km,
corresponding to the Fermilab to DUSEL distance, and consequently, a lower muon
energy at around
$4.5$~GeV~\cite{Geer:2007kn,Bross:2007ts,FernandezMartinez:2010zza}. A detector
with good energy resolution and low energy threshold allows the LENF to exploit
the rich oscillatory pattern and to achieve an important sensitivity to the
oscillation parameters.  In the original proposal, the detector of choice was a
Totally-Active Scintillator Detector (TASD) magnetized by means of a large
magnetic cavern~\cite{Bross:2007ts}.  Other possible detectors include a
magnetized liquid-argon time projection chamber, which would be ideal due to
its large size and excellent detector performance~\cite{Rubbia:1977zz}.  The
possibility of using a non-magnetized detector has also been
studied~\cite{Huber:2008yx}.  In this case, although the event separation
between wrong- and right-sign muons is impossible at an individual event level,
the use of statistical techniques allows the two channels to be partially
distinguished and good sensitivity can be obtained thanks to the high event
numbers which are associated with very large non-magnetized detectors.
Additional studies are required in order to fully understand the capability of
statistical separation of right- and wrong-sign muons and the impact of
backgrounds.  Studies of the LENF using a Magnetized Iron Neutrino Detector
(MIND) have also shown promising performance~\cite{Agarwalla:2010hk} and could
serve as an intermediate step in the development of a future higher-energy
facility. However, this set-up is optimized for large energies due to the poor
low-energy efficiency of the MIND detector and consequently the performance
typically indicates a generic preference for high energies.
A LENF with a TASD has also been considered as part of an
incremental neutrino factory program \cite{Tang:2009wp}. In this setting, the
optimal baseline for a stored-muon energy of $E_\mu=4.12$~GeV was found to be
around $L=1100$~km.

The conventional high-energy neutrino factory (HENF) and the LENF are
conceptually very similar and are only truly separated by a choice of energy
scale.  This suggests that the two designs for the neutrino factory should be
seen as part of a continuum in energy and distance.  
Especially in light of the recent measurement of large $\theta_{13}$,
which has led to the low-energy design becoming the preferred experimental
option,
it is important to perform a ``green-field'' study of the LENF in order to
understand the dependence of the sensitivity to the oscillation parameters on
the stored-muon energy and the choice of baseline distance and, if possible, to
answer the question: {\it what is the optimal LENF?}.  In this paper, we have
performed this study considering ranges for the energy and baseline which
interpolate between the HENF and the LENF and providing an understanding of how
the NF design transitions between the two regions.  
In particular, we have conducted a finely-grained scan over the energy and
baseline range and have identified new performance indicators for two primary
questions: the discovery of CP-violation and of the mass hierarchy. The
analysis that we present has, for the most part, been performed without fixing
the value of $\theta_{13}$. For the measurements of CP-violation and mass
hierarchy this allows us to present results for the full range of possible
values of $\theta_{13}$. For completeness, we have also computed the
discovery potential for measurements of $\theta_{13}$ itself, as the discussion
of this measurement allows the relevant physical effects governing the
performance and sensitivities of the neutrino factory to be explored and
better understood.

After the completion of our work, a study~\cite{Dighe:2011pa} appeared which
also addresses the optimal baseline and stored-muon energy of the LENF.  Our
work differs from \refref{Dighe:2011pa} in a few significant regards. First of
all, we have performed a full, simultaneous scan over the $E_\mu$--$L$
parameter space and as such have a diminished reliance on the interpolation of
results.  Secondly, our study considers three different detector options which
helps us to fully understand the potential performance of the LENF.  Finally,
in this study we have chosen $10^{22}$ total useful muon decays, a range of
baseline distances from $1000$~km to $4000$~km and stored-muon energies from
$4$~GeV to $25$~GeV.  These choices were made to connect previous optimization
studies of the LENF ($1300$~km, $4.5$~GeV) to those of the HENF (short baseline
$\sim4000$~km, $25$~GeV).  Understanding this parameter space in detail allows
the two designs to be viewed as part of a continuum and helps us to unite the
previous work on the two designs in a common framework.

This paper is organized as follows. In \secref{sec:details} we discuss the
details of our simulation and the assumptions we have made in our model of the
LENF. In \secref{sec:results} we present the results of our study and provide
an analysis of the sensitivities of the LENF across our parameter space. 
In \secref{sec:largetheta} we discuss the performance of the LENF in
light of the recently measured value of $\theta_{13}$ and in
\secref{sec:conclusions}, we summarize our findings and make a few concluding
remarks.

\section{\label{sec:details}Simulation Details} 

In this study, we have performed simulations of the physics performance of the
LENF over a range of experimental configurations.  As our indicators of
performance, we have computed the sensitivity of the LENF towards three of the
most important potential discoveries for the future generation of neutrino
oscillation experiments:  the discovery of non-zero $\theta_{13}$, the
discovery of CP-violation arising from the
Pontecorvo-Maki-Nakagawa-Sakata~\cite{Pontecorvo:1957aa, *Pontecorvo:1958aa,
*Mns:1962aa} matrix and the determination of the neutrino mass hierarchy.  We
have studied the potential for discovery of each of these fundamental phenomena
over a range of stored-muon energies given by $4\le E_\mu\le25$~GeV and
baseline distances by $1000\le L \le 4000$~km.  This range connects the regions
of parameter space traditionally associated with the LENF
design~\cite{Geer:2007kn,Bross:2007ts,FernandezMartinez:2010zza} to those of
the conventional HENF set-up~\cite{Choubey:2011zz}.

It has been suggested that $1.4\times10^{21}$ muon decays per year is an
attainable goal for the lower-energy accelerator facility
\cite{Ankenbrandt:2009zza} and this estimate has been incorporated into
previous studies of the performance of the LENF
\cite{FernandezMartinez:2010zza, Li:2010aa}.  In this work, so as to aid
comparison with studies of the standard neutrino factory \cite{Huber:2006wb,
Agarwalla:2010hk}, we have assumed $1.0\times10^{21}$ useful muon decays per
year per polarity and a run-time of $10$ years which is divided evenly between
the two polarities ($10^{22}$ useful muons in total).  This value is in
accordance with the estimates for the conventional neutrino factory, assuming
$10^7$ operational seconds per year \cite{Choubey:2011zz}.  We have
additionally performed simulations assuming the optimized value of
$1.4\times10^{21}$ muon decays per year per polarity, the data indicate a
predictable uniform increase in performance and have been made available
online~\cite{ONLINE}. 
In \refref{Dighe:2011pa}, the total number of useful muon decays was taken as
$2.5\times10^{22}$.  This leads to a further increase in statistics and
consequently in the discovery potential of the LENF. To understand the
performance of the LENF in the context of similar experimental configurations
it is important to ensure a fair comparison is made. For this reason, in the
analysis that follows we have assumed the conventional $10^{22}$ total useful
muon decays. 

The detectors in our simulations measure muons arising through both the
disappearance ($\nu_\mu\to\nu_\mu$ or
$\overline{\nu}_\mu\to\overline{\nu}_\mu$) and the appearance channels
($\overline{\nu}_e\to\overline{\nu}_\mu$ or $\nu_e\to\nu_\mu$).  We have chosen
not to include the platinum channel (the observation of $\nu_e$ and
$\overline{\nu}_e$) as previous work has shown that it offers only marginal
improvement of the oscillation parameter sensitivity
\cite{FernandezMartinez:2010zza}.  The production of $\tau^+$($\tau^-$) in the
detector by the charged-current interactions of incident
$\overline{\nu}_\tau$($\nu_\tau$) leads to the problem of \emph{tau
contamination} \cite{Indumathi:2009hg, Dutta:2011mc}.  $\tau$-leptons have a
lifetime at rest of ${2.9\times10^{-13}}$~s and decay inside the detector into
muons with a branching ratio of around $17\%$ \cite{Nakamura:2010zzi}.  This
effect leads to an increased number of both wrong- and right-sign muons.  Due
to the form of the oscillation probabilities the number of additional muons in
the right-sign channel (e.g. from $\nu_\mu\to\nu_\tau$ for $\mu^-$ in the
storage ring) are considerably larger than in the wrong-sign channel (e.g. from
$\bar{\nu}_e\to\bar{\nu}_\tau$ for $\mu^-$ in the storage ring). Subsequently,
the effect of tau-contamination is most pronounced for measurements which rely
on an accurate determination of the disappearance channels, for example in
investigations of $\theta_{23}$-maximality \cite{Indumathi:2009hg}.
This contamination can lead to serious systematic uncertainties if unaccounted
for. However, it has been shown that correctly incorporating this additional
source of muons into the analysis of the golden channel, via migration
matrices, can resolve the systematic deviations \cite{Donini:2010xk}.  This has
been confirmed in a recent study on the performance of the standard NF
\cite{Agarwalla:2010hk} where the change in sensitivity produced by correct
incorporation of the contamination channel was found to be small. The effect of
tau-contamination is also expected to be smaller for lower-energy facilities as
the number of neutrinos with energies above the tau production threshold will
be reduced. For these reasons, we have omitted the contamination channel in our
study. 

Our simulations were performed numerically using the GLoBES package
\cite{Huber:2004ka,Huber:2007ji} which incorporates the Preliminary Reference
Earth Model \cite{Dziewonski:1981aa,Stacey:1977aa} for the computation of the
matter density along the baseline. In this package, the oscillation probabilities are computed via
numerical diagonalization of the full Hamiltonian assuming three neutrino
flavors.  However, it is convenient during the analysis of our results to
introduce an approximate expression~\cite{Cervera:2000kp, *Cervera:2001zz} for
the golden-channel oscillation probability which is valid up to second-order in
$\theta_{13}$, $\alpha \equiv \Delta m^2_{21} / \Delta m^2_{31}$, $\Delta m^2_{21} L/E$ and $\Delta m^2_{21}/EA$  
\begin{align} P_{e\mu} &= ~\sin^22\theta_{13}\sin^2\theta_{23}\sin^2\left(\frac{\Delta m^2_{31}L}{4E}-\frac{AL}{2}\right)\nonumber \\
&+ \left[ \alpha \sin2\theta_{13}\sin2\theta_{12}\sin2\theta_{23}\frac{\Delta m^2_{31}}{2EA}\sin\left(\frac{AL}{2}\right)\right . \nonumber \\
&\quad~~\left . \times\sin\left( \frac{\Delta m^2_{31}L}{4E}-\frac{AL}{2}\right)\cos\left(\frac{\Delta m^2_{31}L}{4E}+\delta\right)\right]\nonumber \\
&+ \alpha^2\cos^2\theta_{23}\sin^22\theta_{12}\left(\frac{\Delta m^2_{31}}{2EA} \right)^2\sin^2\left(\frac{AL}{2}\right). \label{OscProb}
\end{align}
The first summand in this expression is referred to as the \emph{atmospheric
term} and depends quadratically on $\theta_{13}$. The \emph{CP term} is second
and introduces dependence on $\delta$. The remaining part is called the
\emph{solar term}, which for small values of $\theta_{13}$ can dominate the
oscillation probability. As the solar term is independent of $\theta_{13}$,
$\delta$ and $\text{sign}(\Delta m^2_{31})$, it can lead to a significant loss
of sensitivity in the measurement of these parameters if dominant. 

The optimal detector technology for the LENF has not yet been identified and a
number of candidate designs remain viable.  To facilitate a comparison between
the alternative designs, we have performed our simulations for three detectors
simultaneously: a Totally-Active Scintillator Detector (TASD) and two different
liquid-argon detectors which have optimistic and conservative performance
estimates respectively.  The TASD concept has been successfully implemented in
the MINER$\nu$A experiment \cite{Drakoulakos:2004gn} and a larger scale device
has been selected for the upcoming NO$\nu$A \cite{NOVA2007} design.  Our model
of the TASD is based upon Ref.~\cite{Bross:2007ts} and has a fiducial mass of
$20$~kton, $35$ variable-width energy bins and a constant energy resolution of
$10\%$ for both quasi-elastic and non-quasi-elastic events.  The efficiency
rises linearly from $73\%$ to $94\%$ over a range of $0.5$~GeV to $1$~GeV and
then remains constant for higher energies.  
The background on the golden channel is taken as a constant fraction of
$1\times10^{-3}$ of the events arising from neutral-current interactions and
the same fraction of events from the disappearance channel which accounts for
instances of charge misidentification. 
Both of the liquid-argon detectors 
are based upon parameters first reported in Ref.~\cite{Barger:2007jq} and elaborated on in subsequent optimization
studies \cite{FernandezMartinez:2010zza} 
by having a fiducial mass of $100$~kton, an energy resolution on quasi-elastic
events of $10\%$ and a flat detection efficiency of $80\%$. The conservative
(optimistic) model has $22$ ($35$) variable-width energy bins and an energy
resolution of $20\%$ ($10\%$) on non-quasi-elastic events; the backgrounds are
taken as a fraction of $5\times10^{-3}$ ($1\times10^{-3}$) of events from both
the neutral-current and disappearance channels. 

Although a number of large mass-scale liquid-argon detectors have been
proposed \cite{Bartoszek:2004si, Cline:2006st,
Baibussinov:2007ea,Rubbia:2009md}, only a few designs
\cite{Rubbia:2009md,Cline:2006st} discuss extensions to $100$~kton.  The
magnetization of large-volume detectors, of the scale considered in our
simulations, is a particular challenge and further research is needed to fully
assess the feasibility of the design. Our choice of such large detector volumes
($100$~kton and $20$~kton for liquid-argon and TASD respectively) is designed
to provide an optimistic performance estimate which covers the full range of
potentialities of the LENF if these technical difficulties can be overcome. It
is worth noting that, in our approximation, the performance of the LENF will only depend upon the
detector mass through its \emph{exposure} (number of muon decays $\times$
fiducial detector mass). Using this equivalence, the performance of a
$100$~kton detector with $1.0\times10^{21}$ muon decays per year is expected to
be comparable to a $70$~kton detector with a larger flux of $1.4\times10^{21}$
decays.

We have assumed normal hierarchy to be true throughout our simulations and the known
oscillation parameters were chosen to be $\sin^22\theta_{12}=0.3$,
$\theta_{23}=\pi/4$, $\Delta m^2_{12}=8.0\times10^{-5}~\text{eV}^2$ and
$|\Delta m^2_{13}|=2.5\times10^{-3}~\text{eV}^2$.  The uncertainty on these
values was accounted for by allowing the parameters to vary during the
minimization procedure: we allowed an uncertainty of $4\%$ and $10\%$ for the
solar and the atmospheric parameters, respectively.  These parameter choices
were made in accordance with previous optimization studies of the LENF
\cite{FernandezMartinez:2010zza} and are close to the best-fit values and
uncertainties from recent global analyses of the existing neutrino oscillation
data \cite{GonzalezGarcia:2010er,Schwetz:2011qt}.  

\begin{figure*}
\centering
 \vspace{-0.7cm}
   \subfigure[~TASD: 100\%]{\includegraphics[width=.45\textwidth, viewport=0 0 427 400, clip, angle=0]{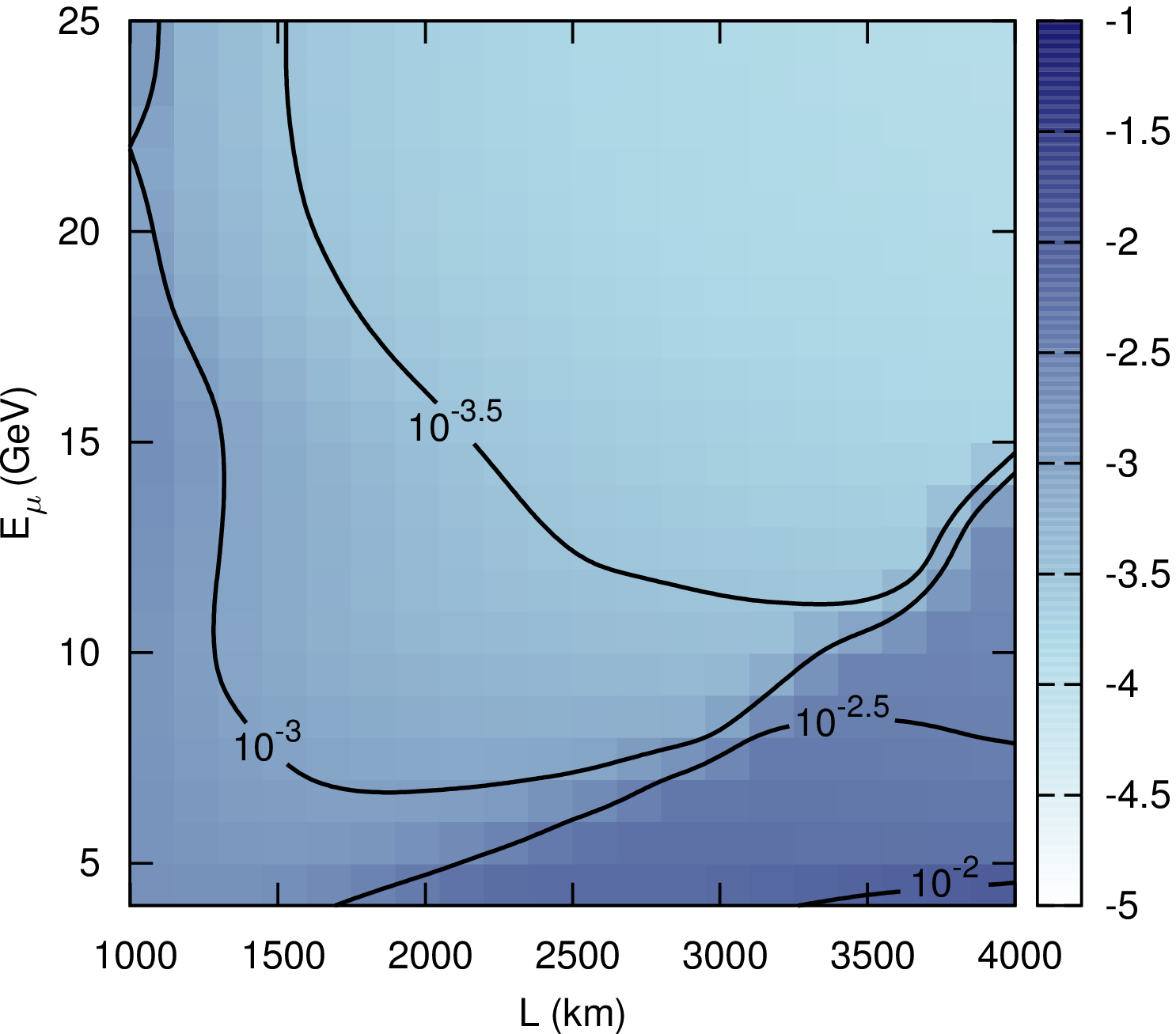}}
 \hspace{0.2cm}
   \subfigure[~TASD: 0\%]{\includegraphics[width=.45\textwidth, viewport=0 0 427 400, clip, angle=0]{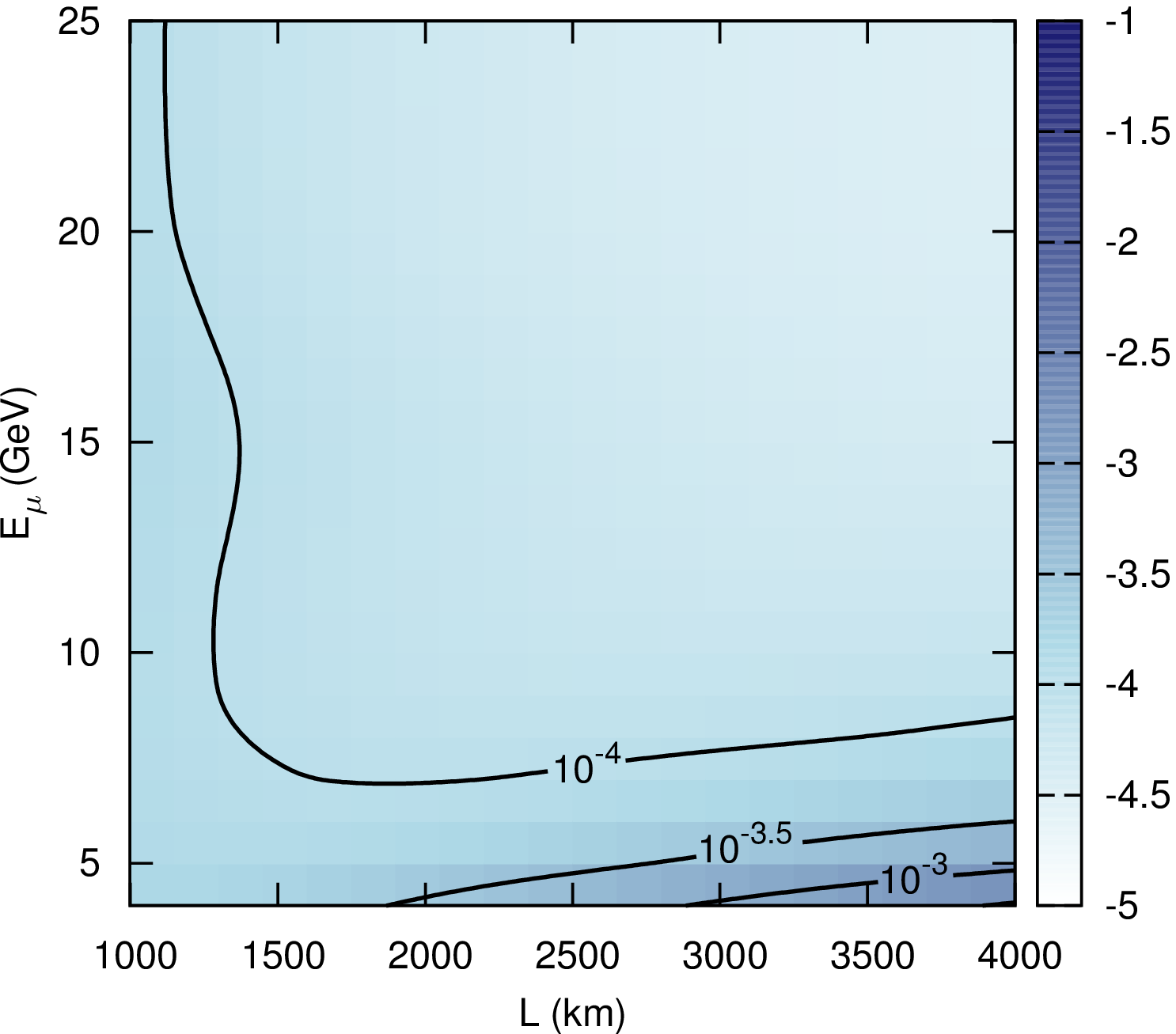}}\\
 \vspace{-0.5cm}
   \subfigure[~Liquid Ar (conservative): 100\%]{\includegraphics[width=.45\textwidth, viewport=0 0 427 400, clip, angle=0]{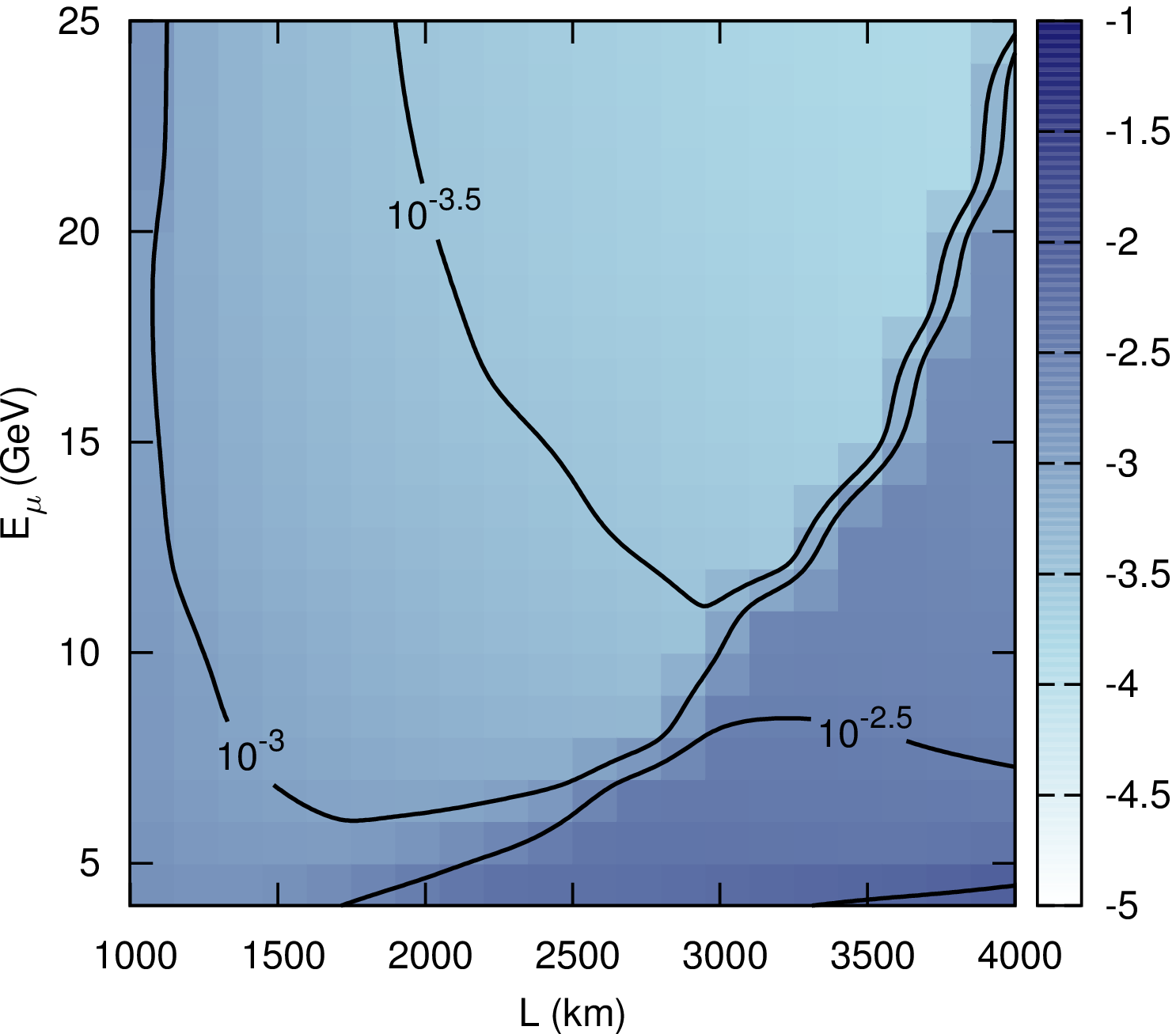}}
 \hspace{0.2cm}
   \subfigure[~Liquid Ar (conservative): 0\%]{\includegraphics[width=.45\textwidth, viewport=0 0 427 400, clip, angle=0]{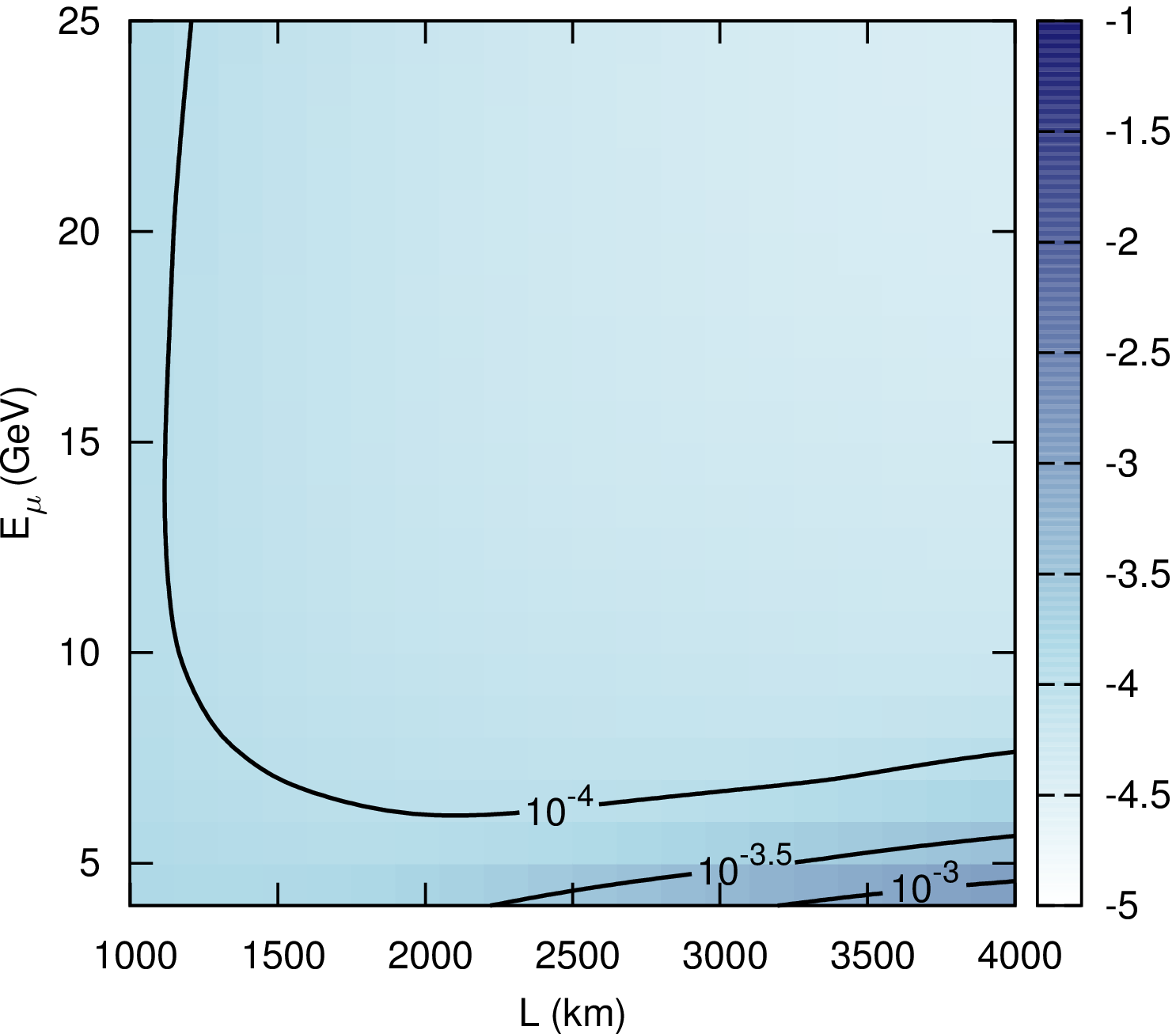}}\\
 \vspace{-0.5cm}
  \subfigure[~Liquid Ar (optimistic): 100\%]{\includegraphics[width=.45\textwidth, viewport=0 0 427 400, clip, angle=0]{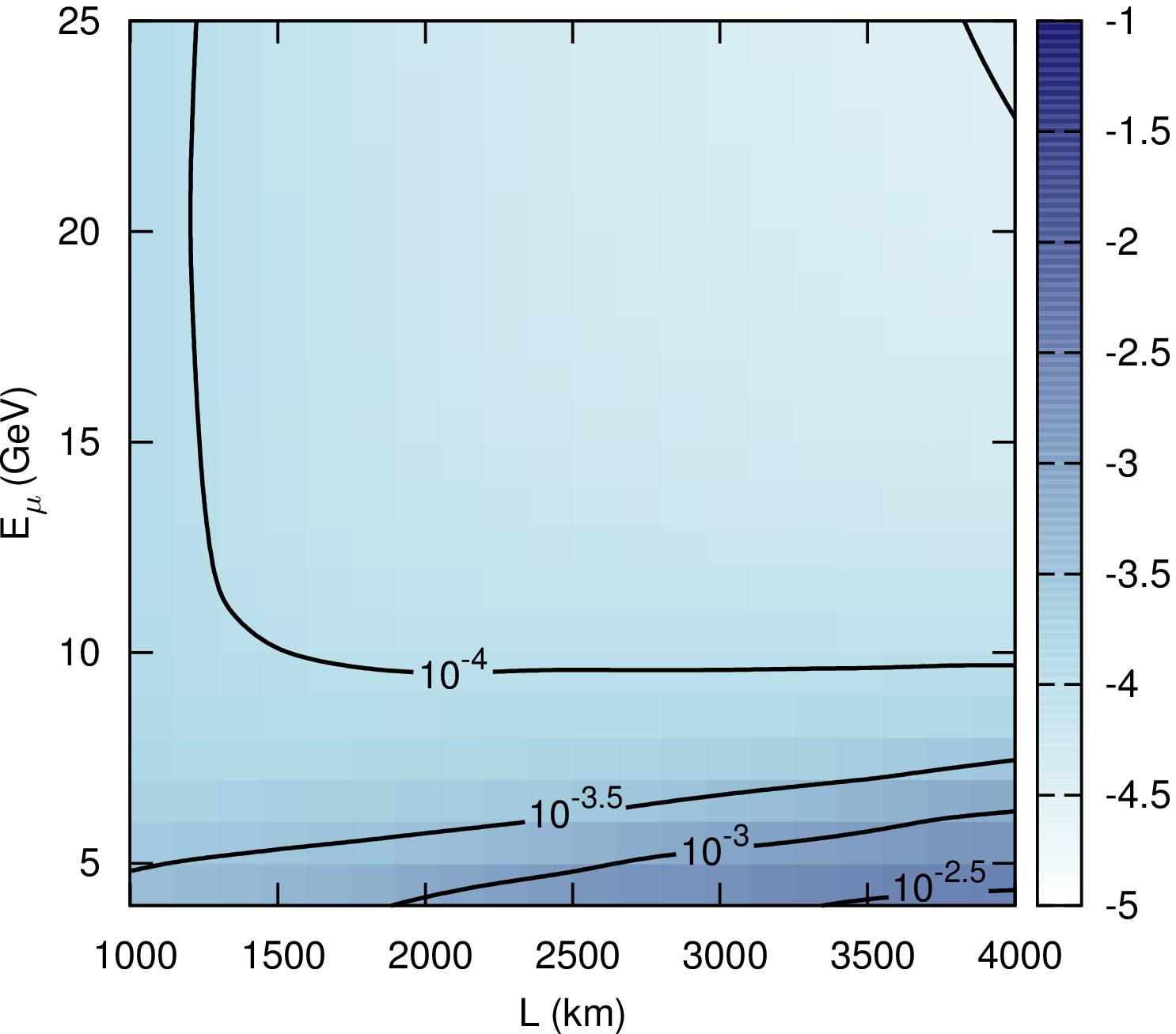}}
 \hspace{0.2cm}
   \subfigure[~Liquid Ar (optimistic): 0\%]{\includegraphics[width=.45\textwidth, viewport=0 0 427 400, clip, angle=0]{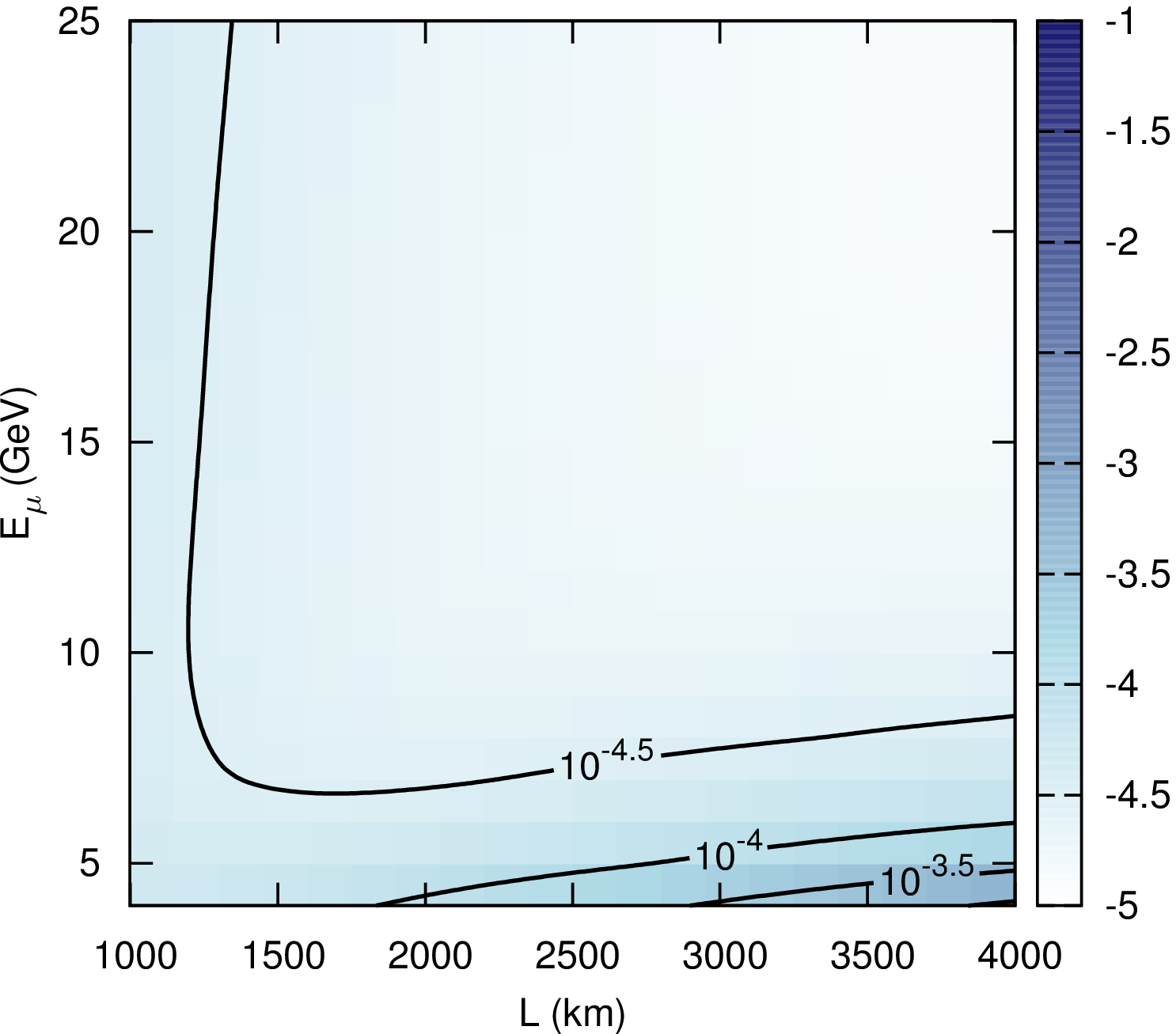}}
\caption{\label{THETA1}The $\theta_{13}$ discovery range as a function of baseline, $L$, and
stored-muon energy, $E_\mu$. The left-hand column shows the lowest value of
$\sin^22\theta_{13}$ for which the discovery fraction is $100\%$ for all higher
values and the right-hand column shows the lowest value of $\sin^22\theta_{13}$
for which there is a non-zero discovery fraction. The intensity of each point is given by the value of $\log_{10}(\sin^22\theta_{13})$.}
\end{figure*}

\section{\label{sec:results}Results and Analysis}

Here we present the performance of the LENF by considering its ability to
make three major discoveries. We start by computing the sensitivity to
non-zero $\theta_{13}$ which, in spite of recent experimental results, is
included to highlight a number of general physical effects which influence the
performance of the the neutrino factory. After this, we focus on the discovery
of $\delta\in\{0,\pi\}$ and $\Delta m^2_{31}>0$. A discussion relevant to the
specific case of large $\theta_{13}$ is presented in \secref{sec:largetheta}.

In this work we compute the discovery potential of non-zero $\theta_{13}$ which
is defined as the exclusion of $\theta_{13}=0$ at the $3\sigma$ confidence
level, marginalizing over all other oscillation parameters.  For the
discovery of CP-violation, we consider the exclusion of all parameter sets with
$\delta\in\{0,\pi\}$ (the CP-conserving values) and for the determination of
the mass hierarchy, we require the exclusion of all parameter sets with the
incorrect hierarchy.  To simplify the presentation of our results we use the
concept of the \emph{discovery fraction}, defined as the fraction of true
values of $\delta$ at which discovery of the related quantity is possible for a
given value of $\theta_{13}$.  For each choice of detector, baseline distance
and stored-muon energy we have computed the discovery fraction as a function of
the true value of $\theta_{13}$ for the three quantities of interest. The full
set of discovery plots computed in this study, showing the regions of
$\sin^22\theta_{13}$--$\delta$ parameter space for which discovery is possible,
has been made available online \cite{ONLINE}. 

To aid our analysis, we will briefly recall some of the generic factors which
influence the variation in performance of the LENF over the $L$--$E_\mu$
parameter space.
Maintaining a high flux of neutrinos is essential in long-baseline experiments
as this determines the number of events that can be observed at the detector
and governs the statistical significance of any observations. 
The total flux incident on the detector increases with the stored-muon
energy as $E^2_\mu$, whilst decreasing with longer baselines as $L^{-2}$.
The observed number of wrong-sign muons is heavily influenced by the
oscillation probability, \equaref{OscProb}, which introduces an additional
dependence on the baseline distance and on the energy of the individual
neutrinos, $E_\nu$.  Observing events which come from the first oscillation
maximum is important as it ensures a large signal in the appearance channel.
Events from the low-energy part of the spectrum contain important
information on CP-violation, 
as for these values of $E_\nu$ the oscillation probability exhibits a strong
dependence on the CP-violating phase $\delta$.
Matter effects lead to an enhancement or suppression of the oscillation signal
compared to the same process in vacuum. Observing the sign of this difference
can provide us with information on the neutrino mass hierarchy. In
\equaref{OscProb}, matter effects are present if $A\neq0$ and their influence
increases with baseline distance and neutrino energy.
Sensitivity to a given measurement can also be affected by the presence of
degeneracies, the locations of which generally depend on both $L$ and $E_\nu$.
At the detector, the neutrino-nucleon interaction cross-sections depend on the
energies of the incident neutrinos. Flavor-tagged detection is made possible
for charged-current interactions by observing the emitted charged lepton. The
total cross-section for this channel increases with
energy~\cite{Lipari:1994pz,Lipari:2002at} and, assuming that all else is kept
equal, leads to an improved number of events at higher-energy facilities. 

To understand the variation in performance of the LENF as we vary $L$ and
$E_\mu$, the individual dependences mentioned above must be considered in
combination. In the following sections, we will discuss how these effects can
explain the sensitivity of the LENF to non-zero $\theta_{13}$, CP violation and the mass
hierarchy.
 
\subsection{\label{subsec:theta}Discovery of non-zero $\theta_{13}$}
 
Here we present the results of our simulation which address the ability of the
LENF to discover non-zero $\theta_{13}$.  
In light of the recent measurements of $\theta_{13}$, optimization towards
this measurement has become largely peripheral. However, the results in this
section, and the accompanying discussion, introduce a number of elements which
govern the performance of the LENF in general. As such, it is instructive to
first understand this measurement before moving on to the question of
CP-violation discovery and the determination of the mass hierarchy. 
A selection of our results can be seen in \figref{THETA1} where the left column
shows, as a function of $L$ and $E_\mu$, 
the lowest value of $\sin^22\theta_{13}$ for which discovery of
$\theta_{13}\neq 0$ can be expected, independently of $\delta$, for all higher
values. This condition is equivalent to ensuring that all higher values of
$\sin^22\theta_{13}$ have discovery fractions of $100\%$.
On the right, \figref{THETA1} shows the lowest value of $\sin^22\theta_{13}$
which can be discovered for at least one value of $\delta$, which corresponds
to a $0\%$ discovery fraction. These can also be seen as the end points of the
$\theta_{13}$ discovery fraction curves seen in previous studies (see for
example Fig.~2 in Ref.~\cite{Agarwalla:2010hk}).  Physically, the first of
these quantities gives the smallest value of $\theta_{13}$ above which we
expect a discovery and provides a conservative estimator of performance.  The
second quantity is a complementary optimistic estimator and tells us the
smallest value of $\theta_{13}$ at which we could possibly make a discovery.
The range of these two parameters gives the region of intermediate performance
where we find discovery fractions between zero and one; discovery in this
region is dependent on the exact value of $\delta$.  

In \figref{THETA1} we see that the TASD expects $\theta_{13}$-discovery to at
least $\sin^22\theta_{13}\gtrsim 10^{-3}$ and has the possibility of extending
this limit by an order of magnitude.  In comparison, the optimistic
liquid-argon detector can discover non-zero $\theta_{13}$ down to at least
$\sin^22\theta_{13}\gtrsim 10^{-4}$ and possibly as low as
$\sin^22\theta_{13}\gtrsim 3\times10^{-5}$.  The TASD and the conservative
liquid-argon detector generally offer comparable sensitivities, which are both
worse than those of the optimistic liquid-argon detector.  For each detector,
we attribute this to different causes. The conservative liquid-argon detector
sees a similar total number of events as the optimistic liquid-argon detector
however its higher backgrounds lead to a poor signal to background ratio which
reduces its sensitivity. In contrast, the TASD has a superior control of the
dominant backgrounds to the conservative liquid-argon detector but sees far
fewer events due to its smaller size. These two effects reduce the attainable
experimental sensitivity by a similar degree.

To further the analysis of our results, we will discuss the four quarters of
this parameter space separately.  In the short-baseline and low-energy region
(SB-LE) with $L\lesssim 2500$~km and $E_\mu\lesssim 14$~GeV, we see $100\%$
discovery fractions for $10^{-4} \lesssim \sin^22\theta_{13} \lesssim
10^{-2.5}$ depending on the choice of detector technology.  
In this region, the atmospheric term is relatively suppressed in the $P_{e\mu}$
oscillation probability. This suppression is a result of small
$\theta_{13}$ and the relative enhancement of the CP and solar terms at lower
neutrino energies. This leads to a poor performance as the signal becomes
decreasingly sensitive to $\theta_{13}$ and the CP term introduces a more
complicated dependence on the oscillation parameters. The decline in
performance found towards the very shortest baselines is a consequence of the
reduction in oscillation probability at small $L$. This leads to a poor
signal-to-background ratio as the number of wrong-sign muons decreases.
In the region of short baselines and high energies (SB-HE) with $L\lesssim
2500$~km and $E_\mu\gtrsim 14$~GeV, the larger energies increase the relative
importance of the atmospheric term and we see that the situation is marginally
improved with respect to the SB-LE region with $100\%$ discovery fractions for
$10^{-4}\lesssim\sin^22\theta_{13}\lesssim 10^{-3}$ depending on detector
technology. The overall improvement in this region can be seen as simply a
leveling-off of the higher energy improvements of the SB-LE regions: for a
fixed baseline distance, there is negligible improvement to be found when
moving from $E_\mu = 15$~GeV to $E_\mu=25$~GeV.  This plateauing effect is
associated with the energy spectrum of the neutrino factory, which rises almost
linearly from $E_\nu=0$ to its peak at $E_\nu\approx2E_\mu/3$ and then drops
sharply at the kinematic cut-off $E_\nu=E_\mu$.  As we increase the stored-muon
energy, the approximately linear tail of this spectrum only decreases slightly
and therefore increasing the stored-muon energy at a LENF can be thought of as
simply adding higher-energy neutrino events on top of the previous low-energy
spectrum. Consequently, the difference in performance between an experiment
with a low and a high stored-muon energy can be estimated by determining the
importance of the neutrino events occurring with energies between the two
stored-muon energies.  Moving up through the SB-HE region, the information
provided by the additional high-energy neutrinos is decreasingly useful because
for higher-energy neutrinos, the $L/E$ ratio is smaller and the oscillation
probability decreases. This effect leads to a law of diminishing returns, where
the discovery reach remains approximately constant. However, as we move to
higher energies, the signal to background ratio decreases and with the
necessary introduction of additional backgrounds and the slight decrease in the
low-energy part of the spectrum, the sensitivity in this region is expected to
ultimately be reduced. 

For the regions of parameter space with baselines of $L\gtrsim 2500$~km, we
again divide the parameter space along $E_\mu=14$~GeV into two quarters: the
long-baseline, low-energy region (LB-LE) and the long-baseline, high-energy
region (LB-HE). The LB-HE region has in general the best $100\%$ discovery
fractions of all of the parameter space.  In this region, the design is
approaching the HENF configuration where the signal to background ratio is
increased and, thanks to the energy-dependent relative suppression of the solar
and CP terms, the atmospheric contribution to the oscillation probability has a
significant influence on the appearance channel.  This permits the true value
of $\theta_{13}$ to be smaller than in the SB-LE region whilst still providing
an appreciable signal and therefore furthering the discovery reach.  In
contrast, the LB-LE region displays the poorest performance across the
parameter space.  We find $100\%$ discovery fractions at around
$\sin^22\theta_{13}\approx 10^{-2.5}$ for the TASD and conservative
liquid-argon detectors, whilst the optimistic liquid-argon detector has a
$100\%$ discovery fraction for $10^{-3}\lesssim
\sin^22\theta_{13}\lesssim10^{-4}$. The poor performance in this region can be
explained as an effect of low statistics: the neutrino flux is reduced as the
baseline increases due to the effects of dispersion on the beam.  In the LB-LE
region this effect is compounded with a small neutrino flux at production.
Consequently, the number of wrong-sign muons incident on the detector becomes
increasingly restricted for these parameter choices.  

In general, the optimal configuration depends upon the magnitude of the true
value of $\theta_{13}$.  
For large values of $\theta_{13}$, as recently confirmed by Daya Bay and RENO\cite{An:2012eh,Ahn:2012nd}, all
configurations are equally capable of confirming this effect.

The $\theta_{13}$ discovery reaches reported here are consistent with studies
performed on the conventional NF \cite{Agarwalla:2010hk} and, although making a
direct quantitative comparison is difficult, the discovery reaches are of a
similar magnitude. This behavior is also reported by \refref{Dighe:2011pa}
although, due to a difference in exposure, once again the results differ
quantitatively.

\begin{figure*}
\centering
\includegraphics[width=16cm]{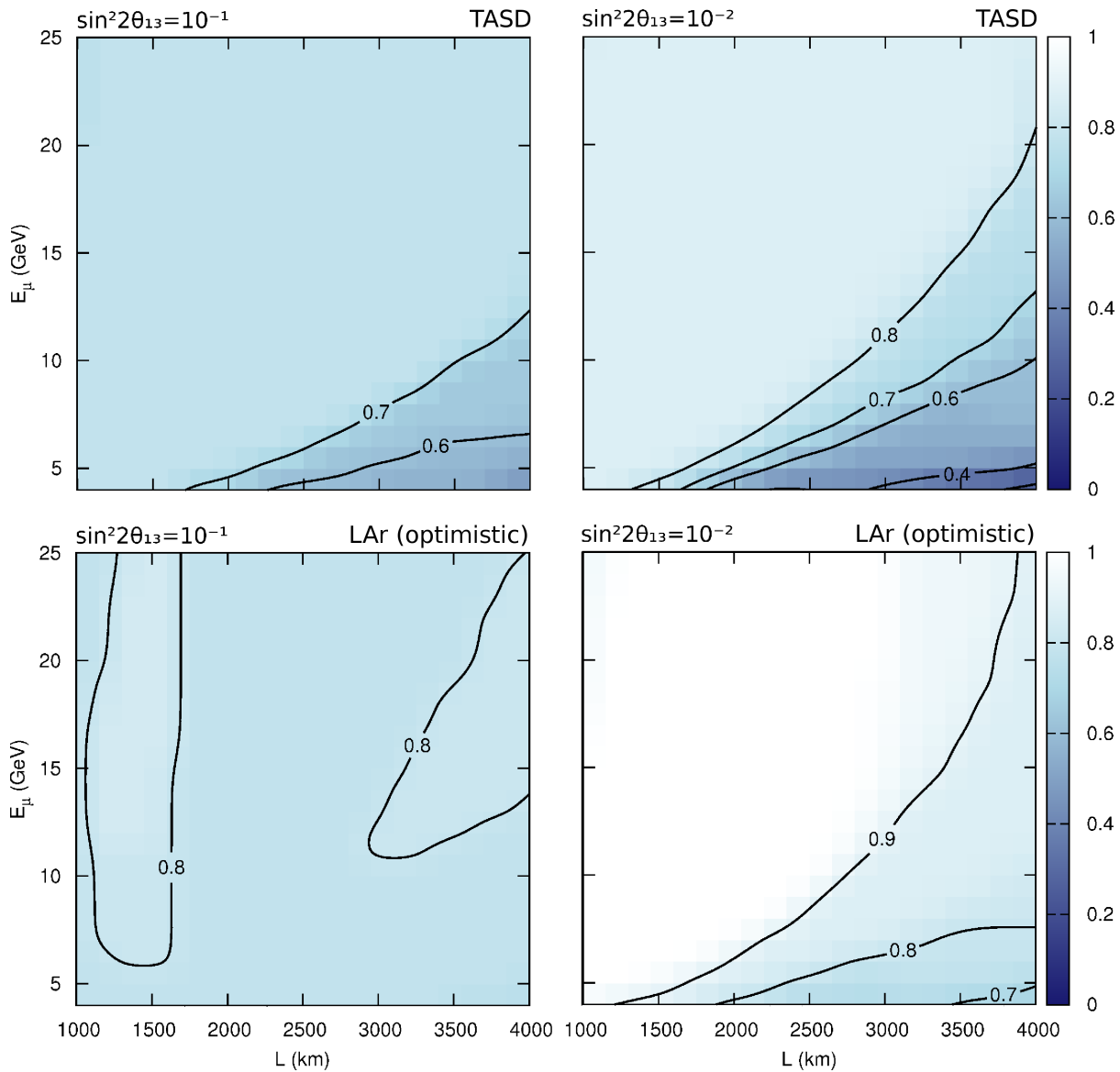}
\caption{\label{CP-all} CP-violation discovery fractions as a function of baseline, $L$, and stored-muon energy, $E_\mu$.  Each column (row) shows the discovery fraction for a different `true' value of $\sin^22\theta_{13}$ (detector option) which is given at the top-left (top-right) of each plot.}
\end{figure*}

\begin{figure*}
\centering
\includegraphics[width=16cm]{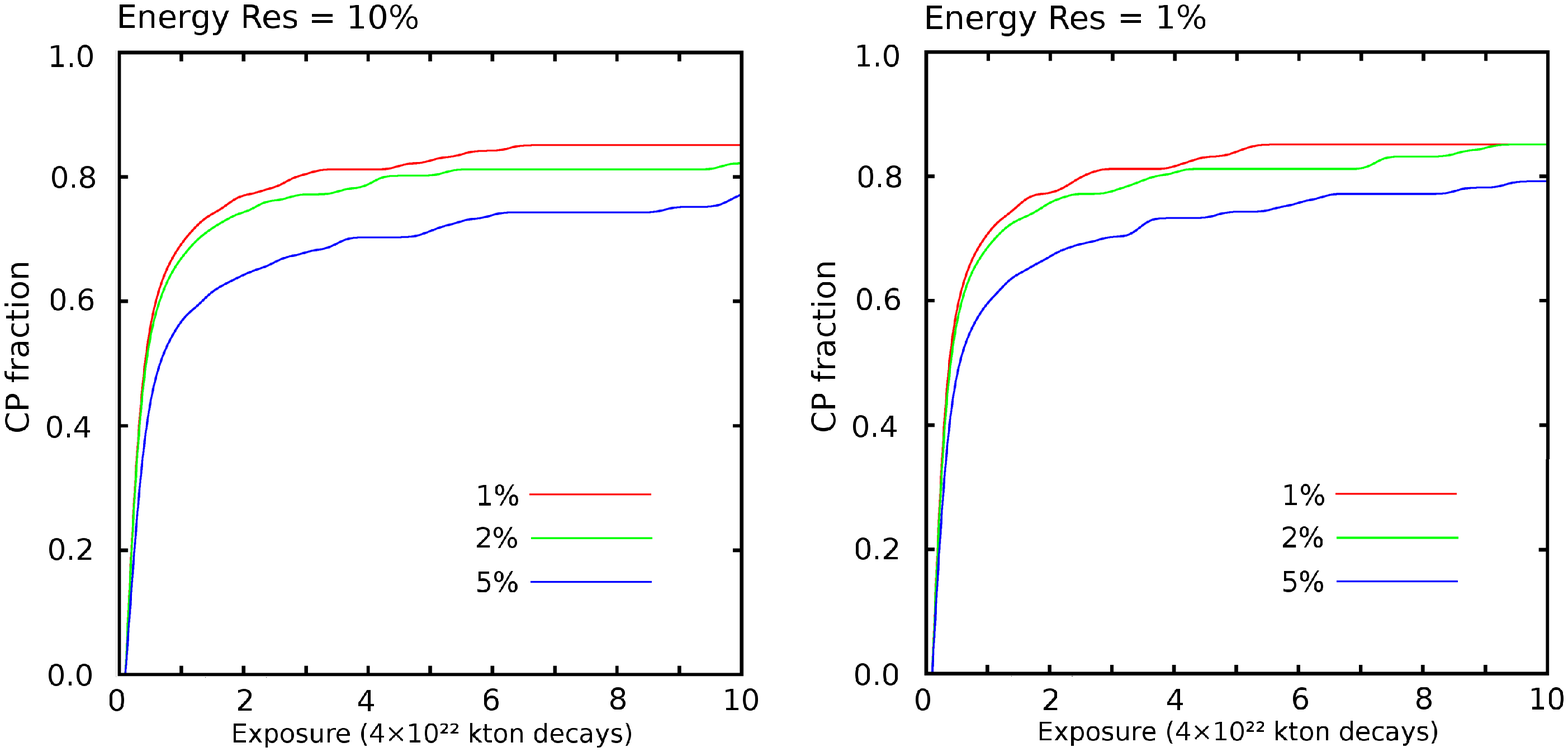}
\caption{\label{CP-exposure} 
The CP-violation discovery fraction as a function of exposure for a range of
systematic errors. The plot on the left (right) shows the discovery fraction
assuming an energy resolution of $10\%$ ($1\%$). The scale of the horizontal
axis is chosen to coincide with runtime assuming a $20$~kton detector and
$10^{21}$ useful muon decays per year per polarity. These plots use the TASD
and have $\theta_{13}$ set to the Daya Bay best-fit value of
$\sin^22\theta_{13}=0.092$~\cite{An:2012eh}. All otherwise unspecified parameters
are as described in \secref{sec:details}.}

\end{figure*}

\subsection{\label{subsec:CP}Discovery of CP-violation}

In \figref{CP-all} we present the CP-violation discovery fraction as a function
of baseline and stored-muon energy at different true values of
$\sin^22\theta_{13}$ for the TASD and the optimistic liquid-argon detector.
The performance of the conservative liquid-argon detector is similar to that of
the TASD and has been omitted from this paper (however, these plots may still
be found online \cite{ONLINE}).  
In the SB-LE region, for $\sin^22\theta_{13}\gtrsim 10^{-2}$ we generally see
very strong performance with discovery fractions of $70\%$ to $90\%$ depending
on the choice of detector technology. These sensitivities confirm the
expectations behind the original motivation for the LENF: a facility with the
ability to measure low-energy neutrino events has access to the oscillation
spectrum near the second maximum where CP-violating effects are most
pronounced. 
The simulation has also been performed for smaller values of
$\sin^22\theta_{13}$ 
where the performance of the SB-LE region starts to worsen as the
$\delta$-dependent terms in the probability are increasingly suppressed. At the
lowest energies, this suppression is compounded by the enhancement of the
$\theta_{13}$- and $\delta$-independent solar term. 
For $\sin^22\theta_{13}\approx10^{-4}$ there is only negligible coverage for
the TASD however the optimistic liquid-argon detector still maintains a
discovery fraction of around $60\%$ due to its combination of high statistics
and strong signal to background ratio. Considering experiments with higher
stored-muon energies, we see scant improvement in the SB-HE region as the
additional events at high energy provide little information on the parts of the
oscillation spectrum which exhibit the most sensitivity to the CP-violating
phase. Generally we see discovery fractions of around $70\%$ to $90\%$ for
$\sin^22\theta_{13}\gtrsim10^{-2}$, which drops to $60\%$ to $70\%$ for
$\sin^22\theta_{13}\approx10^{-3}$. For the smallest values of $\theta_{13}$,
the TASD has once again negligible sensitivity whilst the optimistic
liquid-argon detector can still determine the effect of CP-violation in $60\%$
to $70\%$ of cases.  

As in \secref{subsec:theta}, while the baseline distance is increased there is
a decrease in event numbers due to a weakening of the neutrino flux arising
from long baselines and this may be compounded by an additional weakening of
the flux at low energies.  We generally find the lowest discovery fractions in
the LB-LE region and this contrast is especially marked in the case of small
$\theta_{13}$. However, for long baselines but high energies (LB-HE) we see
good sensitivity to CP-violation, especially when $\theta_{13}$ is small.  In
this region, both the neutrino flux at production and the neutrino-nucleon
cross-sections are increased and this helps to mitigate the effect of baseline
distance on the event numbers.  The additional influence of appreciable matter
effects over longer baselines and the inclusion of neutrinos which probe the
most CP-sensitive parts of the oscillation spectrum further improve the
sensitivity. For $\sin^22\theta_{13}\gtrsim10^{-3}$, the LB-HE region has
comparable discovery fractions to those of the SB-LE region however for
$\sin^22\theta_{13}\approx 10^{-4}$ the only significant sensitivity is to be
found in the LB-HE region with discovery fractions of $40\%$ to $70\%$ for the
TASD and optimistic liquid-argon detectors, respectively. 

As mentioned previously, the true dependence of our simulations on
detector mass, runtime and the number of useful muon decays per year is through
their product, referred to as the \emph{exposure}. In \figref{CP-exposure}, we
can further our understanding of the CP-violation discovery fraction by
considering its dependence on this parameter. In general, for large values of
$\theta_{13}$ the discovery fraction reaches a plateau for each experimental
configuration. This limiting value represents the inherent limitations of the
experiment: exposure directly determines the statistics of our experiment and
at some point the measurement uncertainties will be dominated by systematic
effects for which an increase in statistics can confer only slight improvement
to the experiment's sensitivity.  The plots in \figref{CP-exposure} show the
difference in performance for a modest ($10\%$, left plot) and a more
optimistic ($1\%$, right plot) energy resolution over a range of values of a
uniform systematic error on the signal and backgrounds. We see that variations
in magnitude of the systematic errors induce the greatest change in the
attainable CP-violation discovery fraction.  These effects can lead to a
significant decline in performance; for $10\%$ ($1\%$) energy resolution, there
is a decrease in the discovery fraction of $8\%$ ($6\%$). This limiting
influence on the discovery fraction which arises through systematic
uncertainties is quite stable to variations under baseline and energy, assuming
that these choices do not generate a significant probabilistic suppression of
the number of events. This effect helps to explain the observed uniformity in
parts of \figref{CP-all}, for example in the SB-HE region.
We have also investigated the impact of alternative sources of systematic
limitations. For instance, reducing the prior uncertainty associated with the matter density
can lead to modest improvements in sensitivity; however, 
we find that the unilateral improvement of any one systematic factor leads to
little impact on the sensitivity obtained in \figref{CP-exposure}: it is
necessary to reduce all systematics uniformly to significantly further the
physics reach. Generally, for the parameter ranges that we have studied, it is
the energy resolution and overall systematic error that are responsible for the
greatest variation in the attainable CP-discovery fraction. 

In common with the discovery of $\theta_{13}\neq0$, the optimal configuration of the
LENF for CP-violation discovery divides into two scenarios depending on the
size of $\theta_{13}$.  For $\sin^22\theta_{13}\gtrsim 10^{-3}$, provided the
LB-LE region is avoided, the CP-violation discovery fractions are almost uniform and the
exact configuration is unimportant.  As $\theta_{13}$ decreases, the
$\delta$-dependent signal is suppressed and appreciable sensitivities can only
be found in the LB-HE region where the advantages of the HENF design start to
become relevant.  

The performance of the single-baseline HENF has been
shown~\cite{Agarwalla:2010hk} to share generic features with our data: the
SB-LE region suffers from lower discovery fractions and once $\theta_{13}$ has
decreased to around $\sin^22\theta_{13}=10^{-4}$, the only sensitivity can be
found in the LB-HE region.  The CP-violation discovery fractions of the
single-baseline HENF are very similar to those of the LENF with the TASD
whereas, for all values of $\sin^22\theta_{13}$ that we have studied, the
liquid-argon detector has discovery fractions higher by around $10\%$.  For a
two-baseline HENF, with the longer baseline chosen to be at the magic
baseline~\cite{Agarwalla:2010hk}, the comparison changes depending on the size
of $\sin^22\theta_{13}$.  For $\sin^22\theta_{13}\gtrsim10^{-2}$ the HENF
performs similarly to the LENF with TASD whilst the optimistic liquid-argon
detector has higher discovery fractions, once again by around $10\%$.  For
values of $\theta_{13}$ in the range
$10^{-4}\lesssim\sin^22\theta_{13}\lesssim10^{-3}$, the HENF starts to
out-perform the LENF with TASD.  However, the performance of the LENF with an
optimistic liquid-argon detector remains comparable. Our results in the
SB-LE region agree qualitatively with those computed in
Ref.~\cite{FernandezMartinez:2010zza} for a LENF with $L=1300$~km and
$E_\mu=4.5$~GeV and also agree qualitatively with the recent simulations of the
LENF~\cite{Dighe:2011pa}.

\begin{figure*}
\centering
\includegraphics[width=0.9\textwidth, angle=0]{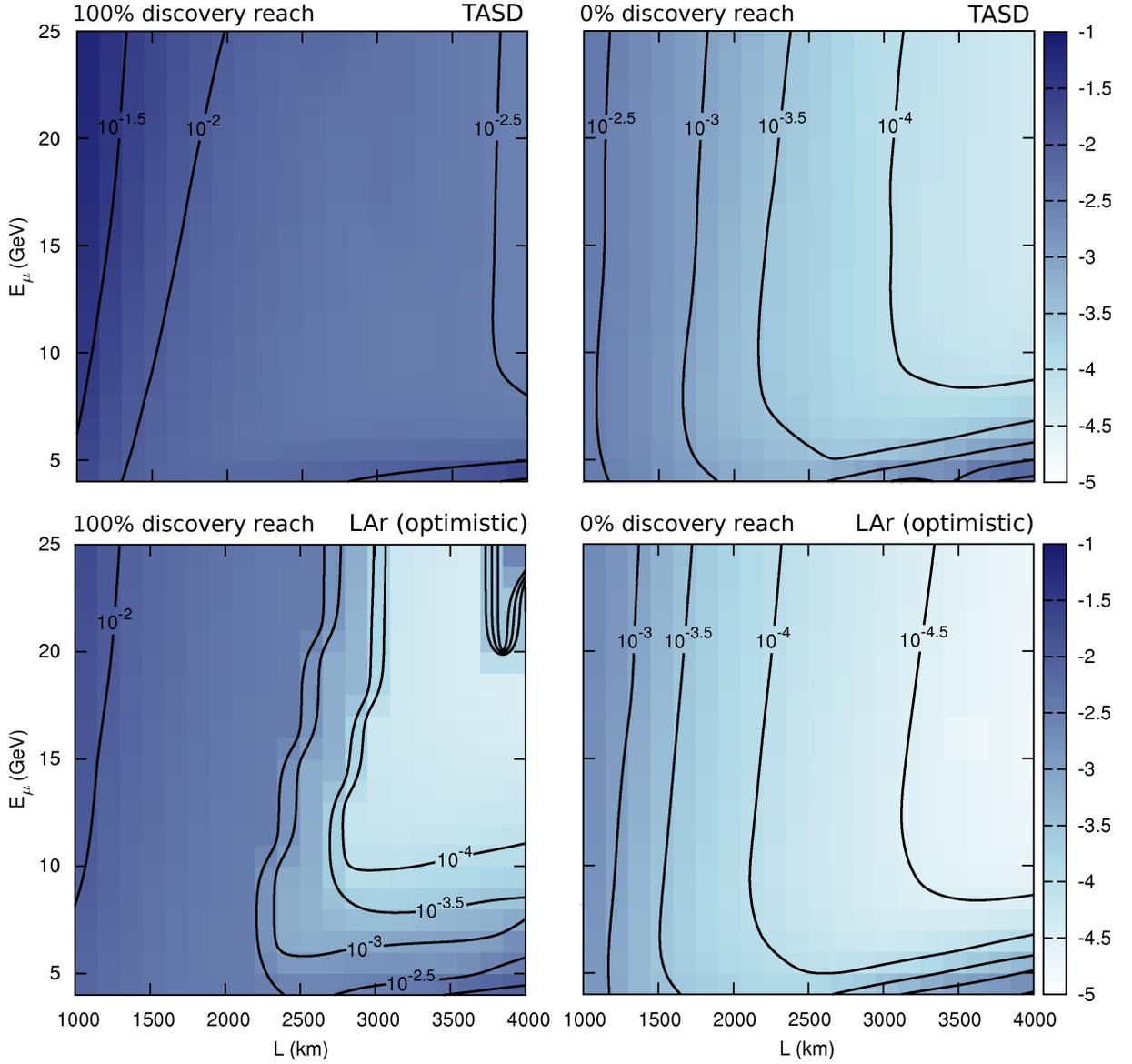}

\caption{\label{HIER}%
Hierarchy determination as a function of baseline, $L$, and stored muon
energy, $E_\mu$, for the TASD (top row) and the liquid-Argon detector with
optimistic performance estimates (bottom row).
The left column shows the most conservative discovery limit: the lowest value of
$\sin^22\theta_{13}$ for which all higher values allow determination of the
mass hierarchy independently of $\delta$. The right column shows the
optimistic limit: the lowest value of $\sin^22\theta_{13}$ for which
there is a non-zero discovery fraction for the exclusion of the incorrect
hierarchy. 
In each plot, the black lines describe regions for which discovery is possible
up to the value of $\sin^22\theta_{13}$ as labeled.
The intensity at each point is given by $\log_{10}(\sin^22\theta_{13})$ of the
relevant discovery limit. }
\end{figure*}

\subsection{Determination of the Mass Hierarchy}

In order to present the results of our simulations regarding the determination
of the mass hierarchy, we have computed analogous quantities to those used in
\secref{subsec:theta}.  In \figref{HIER} we have plotted, as a function of
$L$ and $E_\mu$, the smallest value of $\sin^22\theta_{13}$ at which hierarchy
determination is expected to be possible: the $0\%$ discovery reach. In
\figref{HIER} we show the smallest value of $\sin^22\theta_{13}$ for which
the mass hierarchy can be determined for all higher values of
$\sin^22\theta_{13}$, independently of the value of $\delta$: the $100\%$
discovery reach. We see that, for most of the parameter space, the TASD
provides $100\%$ discovery reaches of around $\sin^22\theta_{13}\gtrsim10^{-2}$
and $0\%$ discovery reaches that are smaller by an order of magnitude.  The
optimistic liquid-argon detector offers a similar $100\%$ discovery reach for
baselines below $L=2500$~km but, for baselines greater than this, can produce a
significantly lower limit between $\sin^22\theta_{13}\gtrsim10^{-3}$ and
$\sin^22\theta_{13}\gtrsim10^{-4}$.  The $0\%$ discovery reach in this case is
generally $\sin^22\theta_{13}\gtrsim3.2\times 10^{-4}$ going down to
$\sin^22\theta_{13}\gtrsim3.2\times10^{-5}$ for the longest baselines at around
$L=3500$~km.   It is clear that these plots exhibit a stronger dependence on
baseline distance than has been seen in previous plots.  This was to be
expected as it is well known that matter effects are crucial in lifting the
hierarchy degeneracy and that these effects increase with longer baselines.
This can be seen in the golden-channel oscillation probability by considering
the difference between the appearance probability for wrong- and right-sign
muons.  Depending on the neutrino mass hierarchy, one of these channels is
suppressed and the other enhanced: this discrepancy grows with longer
baselines. An exception to this pattern is found in the LB-LE region where
there is a notably poor performance compared to the other regions: the
particularly low neutrino flux arising from the combination of long baselines
and low stored-muon energies leads to this decreased sensitivity. These results are in
qualitative agreement with predictions for the standard NF
\cite{Agarwalla:2010hk} and with previous studies of the
LENF~\cite{FernandezMartinez:2010zza,Dighe:2011pa}.

\begin{figure*}
\centering
   \includegraphics[width=.6\textwidth, viewport=10 10 355 390, clip, angle=270]{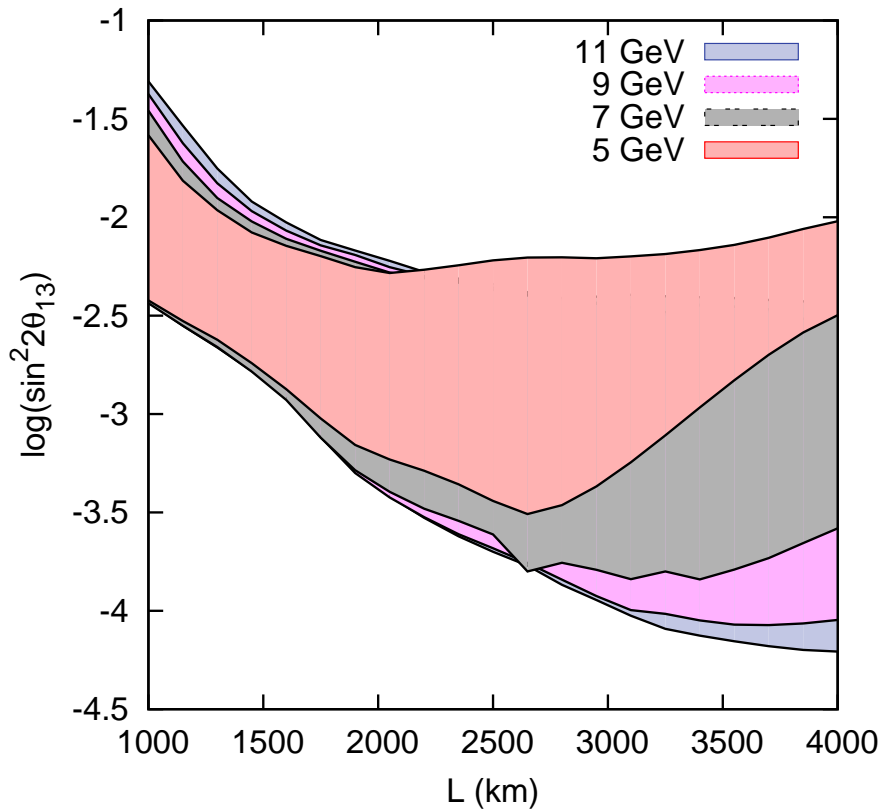}
\caption{The hierarchy discovery reach as a function of baseline distance $L$
for the TASD design. For each band the upper (lower) boundary shows the $100\%$
($0\%$) discovery reach. These correspond to the smallest value of
$\log_{10}(\sin^22\theta_{13})$ for which the hierarchy can be resolved, independently of
$\delta$, for all higher values and the smallest value for which the discovery
can be resolved for at least one value of $\delta$, respectively.}
\label{HIER2}
\end{figure*}

It has been suggested \cite{Dighe:2010js} that a low-energy neutrino factory
with a ``bimagic'' baseline of around $L=2540$~km and stored-muon energy of
$E_\mu=5$~GeV would offer a pronounced sensitivity to the neutrino mass
hierarchy.  This claim was motivated by studies of superbeams
\cite{Raut:2009jj,Joglekar:2010iu} which looked for points of $L$-$E_\mu$
parameter space which show a clean separation between the hierarchy-conjugate
oscillation probabilities.  Such a point was found at $L=2540$~km and
$E=3.3$~GeV where the oscillation probability for the inverted hierarchy is
small and independent of $\delta$.  In contrast, the oscillation probability
for normal hierarchy is much larger and consequently the neutrino fluxes for
different mass hierarchies are expected to differ significantly at this point.
The same property is also found at $L=2540$~km and $E=1.9$~GeV but this time
the probability for the normal hierarchy is small and $\delta$-invariant
whilst that of the inverted hierarchy is large.  
It was the existence of two ``magic'' energies for the same value of $L$ that
lead to the idea of a ``bimagic'' baseline.  A neutrino factory with a
stored-muon energy of $5$~GeV and a baseline of $L=2540$~km would produce a
spectrum covering both of these points and this has been shown to lead to a
strong sensitivity to the neutrino hierarchy \cite{Dighe:2010js}.  In
\figref{HIER2} we show our results indicating how the discovery reach depends
on baseline distance for a selection of muon energies similar to the bimagic
set-up.  At a muon storage energy of $E_\mu=5$~GeV we see evidence for a
minimum in the $0\%$ discovery limit at baseline distances around $2500$ to
$2800$~km.  This lower limit corresponds to the optimistic performance of the
LENF and it is important to note that there is no corresponding minimum in the
conservative estimate shown by the the $100\%$ discovery limit.  We see from
the other bands in \figref{HIER2} how this feature changes as the stored-muon
energy increases: the minimum flattens out and drifts to higher baselines.  We
see that, for all stored-muon energies, a facility with a baseline distance
below $L=2500$~km can improve its discovery reach notably by increasing its
baseline to at least $2500$~km.  Beyond this, it appears that if muon storage
energies higher than $5$~GeV are available then the bimagic choice is not the
optimal configuration as the discovery reach can be additionally furthered by
increasing both the energy and baseline.  Furthermore, it is important to
remember that even if stored muon energies are fixed at $5$~GeV, a baseline of
$L\approx2500$~km only maximizes the optimistic performance of the design and
any potential sensitivity advantage would crucially depend on the value of
$\delta$.  

The behavior in \figref{HIER2} can be understood in light of our previous
analysis. Matter effects are necessary to lift the hierarchy degeneracy and
these are increased significantly by the use of long baselines. The reason that
this trend does not appear for the $E_\mu=5$~GeV case is because, at such a low
energy, the experiment becomes statistically limited in the LB-LE region and
necessitates the move to higher energies. The fact that the configurations near
the bimagic baseline do not confer equal improvements for the conservative,
$100\%$ discovery limit is a result of the construction of the bimagic baseline
criteria: the degree of contrast between hierarchy-conjugate probabilities is
not ensured to be large and its magnitude depends on the exact value of
$\delta$.  Although for some values the contrast is sizable, for others it is
greatly reduced. For example, at the bimagic configuration, when $\delta=0$ the
lowest magic energy at $E_\nu=1.9$~GeV predicts rates which are almost
identical for both hierarchies and can offer little discriminatory information.
The presence of values of $\delta$ for which the hierarchy distinction is less
marked leads to a weaker $100\%$ discovery limit, which is designed to measure
exactly this worst case scenario.

Provided the LB-LE region is avoided, identifying the optimal configuration for
measurements of the mass hierarchy reduces to the observation of a simple
correlation between the expected magnitude of $\sin^22\theta_{13}$ and the
baseline distance. For the smallest values of $\theta_{13}$, configurations in
the LB-HE region are necessary.
However, for the larger values recently measured by Daya Bay and RENO, the physics
reach is quite stable and the exact configuration in our parameter space is
unimportant.

\section{\label{sec:largetheta} Discussion}

In light of the latest experimental results \cite{Abe:2011sj,
Adamson:2011qu, DeKerret:2011aa, An:2012eh, Ahn:2012nd}, the optimization of the LENF
for large values of $\theta_{13}$ has become essential. The Daya Bay and RENO experiments can both independently exclude $\theta_{13}=0$ at around $5\sigma$ and suggest an allowed region of around $0.07\leq\sin^22\theta_{13}\leq 0.14$ at $1\sigma$~\cite{An:2012eh,Ahn:2012nd}.
Our simulations show that these values lie in the optimal region for the
performance of the LENF: CP-violation discovery fractions of $80$--$90\%$ are
attainable in most of the parameter space with an optimistic liquid-argon
detector or TASD.  This sensitivity can be understood by the low-energy
enhancement of the CP term in the golden-channel oscillation probability
combined with a relative suppression of the solar term because of the large
value of $\theta_{13}$.  Additionally, the mass hierarchy will be measurable
independently of the choice of $\delta$ for all of the detectors, baselines and
stored-muon energies that we have considered in this study. This capability
arises because of the effect of the matter potential which generates an
enhancement or suppression in the expected number of wrong-sign muons depending
on the neutrino hierarchy.  For a large value of $\theta_{13}$, a significant
number of oscillated events should occur and it is unlikely that the parameter
sensitivity will be statistically limited. As the motivation for detectors with
very large masses is primarily one of statistics, a large value of
$\theta_{13}$ allows a reduction in detector mass whilst maintaining a similar
level of performance. Although statistical errors are less relevant with a
larger mixing angle, systematic errors become increasingly important.  A
careful study of the systematic uncertainties will be an important next step in
the oscillation analysis.  

For a related discussion of the effect of large values of $\theta_{13}$ on the
sensitivities of the LENF, see also \refref{Dighe:2011pa}. 

\section{\label{sec:conclusions}Conclusions}

The Neutrino Factory has previously been studied in two distinct
configurations: the conventional HENF uses $25$~GeV muons and has two baselines
at $4000$~km and $7000$~km.  More recently, the idea of the LENF has been
presented: this uses low-energy muons, typically around $5$--$10$~GeV, and a
unique baseline at $1500$--$2000$~km.
This set-up exploits the rich oscillatory pattern of the appearance
probability and, by mitigating the effect of degeneracies among the unknown
neutrino parameters, can provide an excellent physics reach for larger values
of $\theta_{13}$.  Motivated by recent experimental results, we have performed
a ``green-field" study of the dependence of the performance of the LENF on the
choice of stored-muon energy, $4\leq E_\mu\leq25$~GeV, and baseline distance,
$1000\leq L\leq4000$~km, in order to ultimately identify the configuration of
the optimal LENF. 

In this paper we have presented the results of numerical simulations on the
ability of the LENF to answer three questions: is $\theta_{13}$ non-zero, does
the neutrino mixing matrix give rise to CP-violating phenomena and what is the
correct neutrino mass hierarchy?  We find that non-zero $\theta_{13}$ can be
discovered for almost all values of baseline and stored-muon energy down to
$\sin^22\theta_{13}\gtrsim 10^{-3}$ and we expect limited sensitivity for a
further order of magnitude.  For CP-violation, discovery fractions of $70\%$ to
$90\%$ are expected to be attainable for $\sin^22\theta_{13}\gtrsim 10^{-2}$
and of $50\%$ to $70\%$ for $\sin^22\theta_{13}\gtrsim 10^{-3}$.  We have shown
that, provided that the extremal configurations of our $E_\mu$--$L$ parameter
space are avoided, specifically very short baselines or long baselines paired
with low energies, these performance estimates are quite general and do not
require significant fine-tuning of the baseline or the stored-muon energy. The
neutrino mass hierarchy is expected to be accessible to the LENF for
$\sin^22\theta_{13}\gtrsim 10^{-2}$ with the possibility of discovery generally
extending to $\sin^22\theta_{13}\gtrsim 4\times10^{-4}$.  For the hierarchy
determination, the dependence on baseline is clearly seen and if baselines of
$L>3000$~km are selected, discovery reach could be extended even further to
$\sin^22\theta_{13}\gtrsim 4\times10^{-5}$. 

We have also considered the potential advantages of the bimagic baseline
configuration given by $L=2540$~km and $E_\mu=5$~GeV.  We have shown that for
this choice of parameters, although a performance maximum is present in the
$0\%$ discovery reach, there is no corresponding maximum in the $100\%$
discovery reach. This means that, if considered conservatively, such a
configuration is of limited benefit for a NF.  We have also shown that higher energy
configurations can provide an improved discovery reach and we see a drift in
the optimal configuration towards higher energies and baselines.

Now that $\theta_{13}$ has been measured to be relatively large and
$\sin^22\theta_{13}\approx0.09$, we see that the sensitivity of the LENF is
comparable for the majority of baseline distance and stored-muon energy
arrangements. 
Furthermore, we expect the results not to be statistically limited, implying
that smaller detectors might be considered to provide the required
sensitivities and that a detailed study of the impact of systematic errors is
required.  For these values of $\theta_{13}$, the optimal LENF for the
resolution of the most compelling open questions of long-baseline physics has a
very broad design. This stability of performance shows that the LENF concept is
versatile, with flexibility to accommodate additional design criteria, whilst
also providing a strong sensitivity to the most important physical
quantities.\\

\section*{Acknowledgements}

The authors would like to acknowledge the involvement of Patrick Huber at the
commencement of this study and for his valuable comments throughout.
This work has been undertaken with partial support from the European Community
under the European Commission Framework Programme 7 Design Studies: LAGUNA
(Project Number 212343) and EURONU (Project Number 212372). SP acknowledges the
support of EuCARD (European Coordination for Accelerator Research and
Development), which is co-funded by the European Commission within the
Framework Programme 7 Capacities Specific Programme, under Grant Agreement
number 227579. PB is further supported by a U.K. Science and Technology
Facilities Council (STFC) studentship.

\bibliographystyle{apsrev4-1}
\bibliography{LENF}{}

\begin{thebibliography}{10}%
\makeatletter
\providecommand \@ifxundefined [1]{%
 \ifx #1\undefined \expandafter \@firstoftwo
 \else \expandafter \@secondoftwo
\fi
}%
\providecommand \@ifnum [1]{%
 \ifnum #1\expandafter \@firstoftwo
 \else \expandafter \@secondoftwo
\fi
}%
\providecommand \enquote [1]{``#1''}%
\providecommand \bibnamefont  [1]{#1}%
\providecommand \bibfnamefont [1]{#1}%
\providecommand \citenamefont [1]{#1}%
\providecommand\href[0]{\@sanitize\@href}%
\providecommand\@href[1]{\endgroup\@@startlink{#1}\endgroup\@@href}%
\providecommand\@@href[1]{#1\@@endlink}%
\providecommand \@sanitize [0]{\begingroup\catcode`\&12\catcode`\#12\relax}%
\@ifxundefined \pdfoutput {\@firstoftwo}{%
 \@ifnum{\z@=\pdfoutput}{\@firstoftwo}{\@secondoftwo}%
}{%
 \providecommand\@@startlink[1]{\leavevmode\special{html:<a href="#1">}}%
 \providecommand\@@endlink[0]{\special{html:</a>}}%
}{%
 \providecommand\@@startlink[1]{%
  \leavevmode
  \pdfstartlink
   attr{/Border[0 0 1 ]/H/I/C[0 1 1]}%
   user{/Subtype/Link/A<</Type/Action/S/URI/URI(#1)>>}%
  \relax
 }%
 \providecommand\@@endlink[0]{\pdfendlink}%
}%
\providecommand \url  [0]{\begingroup\@sanitize \@url }%
\providecommand \@url [1]{\endgroup\@href {#1}{\urlprefix}}%
\providecommand \urlprefix [0]{URL }%
\providecommand \Eprint[0]{\href }%
\@ifxundefined \urlstyle {%
  \providecommand \doi [1]{doi:\discretionary{}{}{}#1}%
}{%
  \providecommand \doi [0]{doi:\discretionary{}{}{}\begingroup
  \urlstyle{rm}\Url }%
}%
\providecommand \doibase [0]{http://dx.doi.org/}%
\providecommand \Doi[1]{\href{\doibase#1}}%
\providecommand \bibAnnote [3]{%
  \BibitemShut{#1}%
  \begin{quotation}\noindent
    \textsc{Key:}\ #2\\\textsc{Annotation:}\ #3%
  \end{quotation}%
}%
\providecommand \bibAnnoteFile [2]{%
  \IfFileExists{#2}{\bibAnnote {#1} {#2} {\input{#2}}}{}%
}%
\providecommand \typeout [0]{\immediate \write \m@ne }%
\providecommand \selectlanguage [0]{\@gobble}%
\providecommand \bibinfo [0]{\@secondoftwo}%
\providecommand \bibfield [0]{\@secondoftwo}%
\providecommand \translation [1]{[#1]}%
\providecommand \BibitemOpen[0]{}%
\providecommand \bibitemStop [0]{}%
\providecommand \bibitemNoStop [0]{.\EOS\space}%
\providecommand \EOS [0]{\spacefactor3000\relax}%
\providecommand \BibitemShut [1]{\csname bibitem#1\endcsname}%
\bibitem{Adamson:2011ig}%
  \BibitemOpen
  \bibfield{author}{%
  \bibinfo {author} {\bibfnamefont{P.}~\bibnamefont{Adamson}} \emph{et~al.}
  (\bibinfo {collaboration} {MINOS~Collaboration}),\ }%
  \bibfield{journal}{%
  \Doi{10.1103/PhysRevLett.106.181801}{\bibinfo {journal} {Phys.~Rev.~Lett.}}\
  }%
  \textbf{\bibinfo {volume} {106}},\ \bibinfo {pages} {181801} (\bibinfo {year}
  {2011})%
  \bibAnnoteFile{NoStop}{Adamson:2011ig}%
\bibitem{Abe:2010hy}%
  \BibitemOpen
  \bibfield{author}{%
  \bibinfo {author} {\bibfnamefont{K.}~\bibnamefont{Abe}} \emph{et~al.}
  (\bibinfo {collaboration} {Super-Kamiokande~Collaboration}),\ }%
  \bibfield{journal}{%
  \Doi{10.1103/PhysRevD.83.052010}{\bibinfo {journal} {Phys.~Rev.~D}}\ }%
  \textbf{\bibinfo {volume} {83}},\ \bibinfo {pages} {052010} (\bibinfo {year}
  {2011})%
  \bibAnnoteFile{NoStop}{Abe:2010hy}%
\bibitem{Wendell:2010md}%
  \BibitemOpen
  \bibfield{author}{%
  \bibinfo {author} {\bibfnamefont{R.}~\bibnamefont{Wendell}} \emph{et~al.}
  (\bibinfo {collaboration} {Kamiokande Collaboration}),\ }%
  \bibfield{journal}{%
  \Doi{10.1103/PhysRevD.81.092004}{\bibinfo {journal} {Phys. Rev. D}}\ }%
  \textbf{\bibinfo {volume} {81}},\ \bibinfo {pages} {092004} (\bibinfo {year}
  {2010})%
  \bibAnnoteFile{NoStop}{Wendell:2010md}%
\bibitem{Aharmim:2011vm}%
  \BibitemOpen
  \bibfield{author}{%
  \bibinfo {author} {\bibfnamefont{B.}~\bibnamefont{Aharmim}} \emph{et~al.}
  (\bibinfo {collaboration} {SNO Collaboration})\ }%
  \Eprint{http://arxiv.org/abs/1109.0763}{arXiv:1109.0763 [nucl-ex]}%
  \bibAnnoteFile{NoStop}{Aharmim:2011vm}%
\bibitem{Derbin:2010zz}%
  \BibitemOpen
  \bibfield{author}{%
  \bibinfo {author} {\bibfnamefont{A.~V.}\ \bibnamefont{Derbin}} (\bibinfo
  {collaboration} {Borexino Collaboration}),\ }%
  \bibfield{journal}{%
  \Doi{10.1134/S1063778810110165, 10.1134/S1063778810110165}{\bibinfo {journal}
  {Phys. Atom. Nucl.}}\ }%
  \textbf{\bibinfo {volume} {73}},\ \bibinfo {pages} {1935} (\bibinfo {year}
  {2010})%
  \bibAnnoteFile{NoStop}{Derbin:2010zz}%
\bibitem{Abe:2008ee}%
  \BibitemOpen
  \bibfield{author}{%
  \bibinfo {author} {\bibfnamefont{S.}~\bibnamefont{Abe}} \emph{et~al.}
  (\bibinfo {collaboration} {KamLAND Collaboration}),\ }%
  \bibfield{journal}{%
  \Doi{10.1103/PhysRevLett.100.221803}{\bibinfo {journal} {Phys. Rev. Lett.}}\
  }%
  \textbf{\bibinfo {volume} {100}},\ \bibinfo {pages} {221803} (\bibinfo {year}
  {2008})%
  \bibAnnoteFile{NoStop}{Abe:2008ee}%
\bibitem{GonzalezGarcia:2010er}%
  \BibitemOpen
  \bibfield{author}{%
  \bibinfo {author} {\bibfnamefont{M.~C.}\ \bibnamefont{Gonzalez~Garcia}},
  \bibinfo {author} {\bibfnamefont{M.}~\bibnamefont{Maltoni}},\ and\ \bibinfo
  {author} {\bibfnamefont{J.}~\bibnamefont{Salvado}},\ }%
  \bibfield{journal}{%
  \Doi{10.1007/JHEP04(2010)056}{\bibinfo {journal} {JHEP}}\ }%
  \textbf{\bibinfo {volume} {1004}},\ \bibinfo {pages} {056} (\bibinfo {year}
  {2010})%
  \bibAnnoteFile{NoStop}{GonzalezGarcia:2010er}%
\bibitem{Fogli:2011qn}%
  \BibitemOpen
  \bibfield{author}{%
  \bibinfo {author} {\bibfnamefont{G.~L.}\ \bibnamefont{Fogli}}, \bibinfo
  {author} {\bibfnamefont{E.}~\bibnamefont{Lisi}}, \bibinfo {author}
  {\bibfnamefont{A.}~\bibnamefont{Marrone}}, \bibinfo {author}
  {\bibfnamefont{A.}~\bibnamefont{Palazzo}},\ and\ \bibinfo {author}
  {\bibfnamefont{A.~M.}\ \bibnamefont{Rotunno}},\ }%
  \bibfield{journal}{%
  \Doi{10.1103/PhysRevD.84.053007}{\bibinfo {journal} {Phys. Rev. D}}\ }%
  \textbf{\bibinfo {volume} {84}},\ \bibinfo {pages} {053007} (\bibinfo {year}
  {2011})%
  \bibAnnoteFile{NoStop}{Fogli:2011qn}%
\bibitem{Schwetz:2011qt}%
  \BibitemOpen
  \bibfield{author}{%
  \bibinfo {author} {\bibfnamefont{T.}~\bibnamefont{Schwetz}}, \bibinfo
  {author} {\bibfnamefont{M.}~\bibnamefont{T\'ortola}},\ and\ \bibinfo {author}
  {\bibfnamefont{J.~W.~F.}\ \bibnamefont{Valle}},\ }%
  \bibfield{journal}{%
  \Doi{10.1088/1367-2630/13/6/063004}{\bibinfo {journal} {New J. Phys.}}\ }%
  \textbf{\bibinfo {volume} {13}},\ \bibinfo {pages} {063004} (\bibinfo {year}
  {2011})%
  \bibAnnoteFile{NoStop}{Schwetz:2011qt}%
\bibitem{Schwetz:2011zk}%
  \BibitemOpen
  \bibfield{author}{%
  \bibinfo {author} {\bibfnamefont{T.}~\bibnamefont{Schwetz}}, \bibinfo
  {author} {\bibfnamefont{M.}~\bibnamefont{T\'ortola}},\ and\ \bibinfo {author}
  {\bibfnamefont{J.~W.~F.}\ \bibnamefont{Valle}}\ }%
  \Eprint{http://arxiv.org/abs/1108.1376}{arXiv:1108.1376 [hep-ph]}%
  \bibAnnoteFile{NoStop}{Schwetz:2011zk}%
\bibitem{Abe:2011sj}%
  \BibitemOpen
  \bibfield{author}{%
  \bibinfo {author} {\bibfnamefont{K.}~\bibnamefont{Abe}} \emph{et~al.}
  (\bibinfo {collaboration} {T2K Collaboration}),\ }%
  \bibfield{journal}{%
  \Doi{10.1103/PhysRevLett.107.041801}{\bibinfo {journal} {Phys. Rev. Lett.}}\
  }%
  \textbf{\bibinfo {volume} {107}},\ \bibinfo {pages} {041801} (\bibinfo {year}
  {2011})%
  \bibAnnoteFile{NoStop}{Abe:2011sj}%
\bibitem{Adamson:2011qu}%
  \BibitemOpen
  \bibfield{author}{%
  \bibinfo {author} {\bibfnamefont{P.}~\bibnamefont{Adamson}} \emph{et~al.}
  (\bibinfo {collaboration} {MINOS Collaboration}),\ }%
  \bibfield{journal}{%
  \Doi{10.1103/PhysRevLett.107.181802}{\bibinfo {journal} {Phys. Rev. Lett.}}\
  }%
  \textbf{\bibinfo {volume} {107}},\ \bibinfo {pages} {181802} (\bibinfo {year}
  {2011})%
  \bibAnnoteFile{NoStop}{Adamson:2011qu}%
\bibitem{DeKerret:2011aa}%
  \BibitemOpen
  \bibfield{author}{%
  \bibinfo {author} {\bibfnamefont{H.}~\bibnamefont{de~Kerret}} (\bibinfo
  {collaboration} {{Double Chooz Collaboration}}),\ }%
  \enquote{\bibinfo {title} {{The first results from the Double Chooz
  experiment}},}\ \bibinfo {note} {Talk presented at LowNu11, Seoul.
  9/11/2011}%
  \bibAnnoteFile{NoStop}{DeKerret:2011aa}%
\bibitem{An:2012eh}%
  \BibitemOpen
  \bibfield{author}{%
  \bibinfo {author} {\bibfnamefont{F.}~\bibnamefont{An}} \emph{et~al.}
  (\bibinfo {collaboration} {DAYA-BAY Collaboration})}%
   (\bibinfo {year} {2012}),\ \bibinfo {note} {5 figures. Version to appear in
  Phys. Rev. Lett},\ \Eprint{http://arxiv.org/abs/1203.1669}{arXiv:1203.1669
  [hep-ex]}%
  \bibAnnoteFile{NoStop}{An:2012eh}%
\bibitem{Ahn:2012nd}%
  \BibitemOpen
  \bibfield{author}{%
  \bibinfo {author} {\bibfnamefont{J.}~\bibnamefont{Ahn}} \emph{et~al.}
  (\bibinfo {collaboration} {RENO collaboration})}%
   (\bibinfo {year} {2012}),\
  \Eprint{http://arxiv.org/abs/1204.0626}{arXiv:1204.0626 [hep-ex]}%
  \bibAnnoteFile{NoStop}{Ahn:2012nd}%
\bibitem{Fogli:1996pv}%
  \BibitemOpen
  \bibfield{author}{%
  \bibinfo {author} {\bibfnamefont{G.~L.}\ \bibnamefont{Fogli}}\ and\ \bibinfo
  {author} {\bibfnamefont{E.}~\bibnamefont{Lisi}},\ }%
  \bibfield{journal}{%
  \Doi{10.1103/PhysRevD.54.3667}{\bibinfo {journal} {Phys.Rev.}}\ }%
  \textbf{\bibinfo {volume} {D54}},\ \bibinfo {pages} {3667} (\bibinfo {year}
  {1996})%
  \bibAnnoteFile{NoStop}{Fogli:1996pv}%
\bibitem{BurguetCastell:2001ez}%
  \BibitemOpen
  \bibfield{author}{%
  \bibinfo {author} {\bibfnamefont{J.}~\bibnamefont{Burguet~Castell}}, \bibinfo
  {author} {\bibfnamefont{M.~B.}\ \bibnamefont{Gavela}}, \bibinfo {author}
  {\bibfnamefont{J.~J.}\ \bibnamefont{G\'omez~Cadenas}}, \bibinfo {author}
  {\bibfnamefont{P.}~\bibnamefont{Hern\'andez}},\ and\ \bibinfo {author}
  {\bibfnamefont{O.}~\bibnamefont{Mena}},\ }%
  \bibfield{journal}{%
  \Doi{10.1016/S0550-3213(01)00248-6}{\bibinfo {journal} {Nucl. Phys. B}}\ }%
  \textbf{\bibinfo {volume} {608}},\ \bibinfo {pages} {301} (\bibinfo {year}
  {2001})%
  \bibAnnoteFile{NoStop}{BurguetCastell:2001ez}%
\bibitem{Minakata:2001qm}%
  \BibitemOpen
  \bibfield{author}{%
  \bibinfo {author} {\bibfnamefont{H.}~\bibnamefont{Minakata}}\ and\ \bibinfo
  {author} {\bibfnamefont{H.}~\bibnamefont{Nunokawa}},\ }%
  \bibfield{journal}{%
  \bibinfo {journal} {JHEP}\ }%
  \textbf{\bibinfo {volume} {0110}},\ \bibinfo {pages} {001} (\bibinfo {year}
  {2001})%
  \bibAnnoteFile{NoStop}{Minakata:2001qm}%
\bibitem{Barger:2001yr}%
  \BibitemOpen
  \bibfield{author}{%
  \bibinfo {author} {\bibfnamefont{V.}~\bibnamefont{Barger}}, \bibinfo {author}
  {\bibfnamefont{D.}~\bibnamefont{Marfatia}},\ and\ \bibinfo {author}
  {\bibfnamefont{K.}~\bibnamefont{Whisnant}},\ }%
  \bibfield{journal}{%
  \Doi{10.1103/PhysRevD.65.073023}{\bibinfo {journal} {Phys. Rev. D}}\ }%
  \textbf{\bibinfo {volume} {65}},\ \bibinfo {pages} {073023} (\bibinfo {year}
  {2002})%
  \bibAnnoteFile{NoStop}{Barger:2001yr}%
\bibitem{Huber:2002mx}%
  \BibitemOpen
  \bibfield{author}{%
  \bibinfo {author} {\bibfnamefont{P.}~\bibnamefont{Huber}}, \bibinfo {author}
  {\bibfnamefont{M.}~\bibnamefont{Lindner}},\ and\ \bibinfo {author}
  {\bibfnamefont{W.}~\bibnamefont{Winter}},\ }%
  \bibfield{journal}{%
  \Doi{10.1016/S0550-3213(02)00825-8}{\bibinfo {journal} {Nucl.Phys.}}\ }%
  \textbf{\bibinfo {volume} {B645}},\ \bibinfo {pages} {3} (\bibinfo {year}
  {2002})%
  \bibAnnoteFile{NoStop}{Huber:2002mx}%
\bibitem{Minakata:2002qi}%
  \BibitemOpen
  \bibfield{author}{%
  \bibinfo {author} {\bibfnamefont{H.}~\bibnamefont{Minakata}}, \bibinfo
  {author} {\bibfnamefont{H.}~\bibnamefont{Nunokawa}},\ and\ \bibinfo {author}
  {\bibfnamefont{S.~J.}\ \bibnamefont{Parke}},\ }%
  \bibfield{journal}{%
  \Doi{10.1103/PhysRevD.66.093012}{\bibinfo {journal} {Phys. Rev. D}}\ }%
  \textbf{\bibinfo {volume} {66}},\ \bibinfo {pages} {093012} (\bibinfo {year}
  {2002})%
  \bibAnnoteFile{NoStop}{Minakata:2002qi}%
\bibitem{Donini:2003vz}%
  \BibitemOpen
  \bibfield{author}{%
  \bibinfo {author} {\bibfnamefont{A.}~\bibnamefont{Donini}}, \bibinfo {author}
  {\bibfnamefont{D.}~\bibnamefont{Meloni}},\ and\ \bibinfo {author}
  {\bibfnamefont{S.}~\bibnamefont{Rigolin}},\ }%
  \bibfield{journal}{%
  \Doi{10.1088/1126-6708/2004/06/011}{\bibinfo {journal} {JHEP}}\ }%
  \textbf{\bibinfo {volume} {0406}},\ \bibinfo {pages} {011} (\bibinfo {year}
  {2004})%
  \bibAnnoteFile{NoStop}{Donini:2003vz}%
\bibitem{Aoki:2003kc}%
  \BibitemOpen
  \bibfield{author}{%
  \bibinfo {author} {\bibfnamefont{M.}~\bibnamefont{Aoki}}, \bibinfo {author}
  {\bibfnamefont{K.}~\bibnamefont{Hagiwara}},\ and\ \bibinfo {author}
  {\bibfnamefont{N.}~\bibnamefont{Okamura}},\ }%
  \bibfield{journal}{%
  \Doi{10.1016/j.physletb.2004.12.022}{\bibinfo {journal} {Phys.Lett.}}\ }%
  \textbf{\bibinfo {volume} {B606}},\ \bibinfo {pages} {371} (\bibinfo {year}
  {2005})%
  \bibAnnoteFile{NoStop}{Aoki:2003kc}%
\bibitem{Yasuda:2004gu}%
  \BibitemOpen
  \bibfield{author}{%
  \bibinfo {author} {\bibfnamefont{O.}~\bibnamefont{Yasuda}},\ }%
  \bibfield{journal}{%
  \Doi{10.1088/1367-2630/6/1/083}{\bibinfo {journal} {New J.Phys.}}\ }%
  \textbf{\bibinfo {volume} {6}},\ \bibinfo {pages} {83} (\bibinfo {year}
  {2004})%
  \bibAnnoteFile{NoStop}{Yasuda:2004gu}%
\bibitem{Kajita:2001sb}%
  \BibitemOpen
  \bibfield{author}{%
  \bibinfo {author} {\bibfnamefont{T.}~\bibnamefont{Kajita}}, \bibinfo {author}
  {\bibfnamefont{H.}~\bibnamefont{Minakata}},\ and\ \bibinfo {author}
  {\bibfnamefont{H.}~\bibnamefont{Nunokawa}},\ }%
  \bibfield{journal}{%
  \Doi{10.1016/S0370-2693(02)01231-5}{\bibinfo {journal} {Phys. Lett. B}}\ }%
  \textbf{\bibinfo {volume} {528}},\ \bibinfo {pages} {245} (\bibinfo {year}
  {2002})%
  \bibAnnoteFile{NoStop}{Kajita:2001sb}%
\bibitem{BurguetCastell:2002qx}%
  \BibitemOpen
  \bibfield{author}{%
  \bibinfo {author} {\bibfnamefont{J.}~\bibnamefont{Burguet~Castell}}, \bibinfo
  {author} {\bibfnamefont{M.~B.}\ \bibnamefont{Gavela}}, \bibinfo {author}
  {\bibfnamefont{J.~J.}\ \bibnamefont{G\'omez~Cadenas}}, \bibinfo {author}
  {\bibfnamefont{P.}~\bibnamefont{Hern\'andez}},\ and\ \bibinfo {author}
  {\bibfnamefont{O.}~\bibnamefont{Mena}},\ }%
  \bibfield{journal}{%
  \Doi{10.1016/S0550-3213(02)00872-6}{\bibinfo {journal} {Nucl. Phys. B}}\ }%
  \textbf{\bibinfo {volume} {646}},\ \bibinfo {pages} {301} (\bibinfo {year}
  {2002})%
  \bibAnnoteFile{NoStop}{BurguetCastell:2002qx}%
\bibitem{Minakata:2003ca}%
  \BibitemOpen
  \bibfield{author}{%
  \bibinfo {author} {\bibfnamefont{H.}~\bibnamefont{Minakata}}, \bibinfo
  {author} {\bibfnamefont{H.}~\bibnamefont{Nunokawa}},\ and\ \bibinfo {author}
  {\bibfnamefont{S.~J.}\ \bibnamefont{Parke}},\ }%
  \bibfield{journal}{%
  \Doi{10.1103/PhysRevD.68.013010}{\bibinfo {journal} {Phys. Rev. D}}\ }%
  \textbf{\bibinfo {volume} {68}},\ \bibinfo {pages} {013010} (\bibinfo {year}
  {2003})%
  \bibAnnoteFile{NoStop}{Minakata:2003ca}%
\bibitem{Diwan:2003bp}%
  \BibitemOpen
  \bibfield{author}{%
  \bibinfo {author} {\bibfnamefont{M.}~\bibnamefont{Diwan}}, \bibinfo {author}
  {\bibfnamefont{D.}~\bibnamefont{Beavis}}, \bibinfo {author}
  {\bibfnamefont{M.-C.}\ \bibnamefont{Chen}}, \bibinfo {author}
  {\bibfnamefont{J.}~\bibnamefont{Gallardo}}, \bibinfo {author}
  {\bibfnamefont{S.}~\bibnamefont{Kahn}}, \emph{et~al.},\ }%
  \bibfield{journal}{%
  \Doi{10.1103/PhysRevD.68.012002}{\bibinfo {journal} {Phys.Rev.}}\ }%
  \textbf{\bibinfo {volume} {D68}},\ \bibinfo {pages} {012002} (\bibinfo {year}
  {2003})%
  \bibAnnoteFile{NoStop}{Diwan:2003bp}%
\bibitem{Huber:2010dx}%
  \BibitemOpen
  \bibfield{author}{%
  \bibinfo {author} {\bibfnamefont{P.}~\bibnamefont{Huber}}\ and\ \bibinfo
  {author} {\bibfnamefont{J.}~\bibnamefont{Kopp}},\ }%
  \bibfield{journal}{%
  \Doi{10.1007/JHEP03(2011)013, 10.1007/JHEP05(2011)024,
  10.1007/JHEP03(2011)013, 10.1007/JHEP05(2011)024}{\bibinfo {journal} {JHEP}}\
  }%
  \textbf{\bibinfo {volume} {1103}},\ \bibinfo {pages} {013} (\bibinfo {year}
  {2011})%
  \bibAnnoteFile{NoStop}{Huber:2010dx}%
\bibitem{Barger:1980tf}%
  \BibitemOpen
  \bibfield{author}{%
  \bibinfo {author} {\bibfnamefont{V.~D.}\ \bibnamefont{Barger}}, \bibinfo
  {author} {\bibfnamefont{K.}~\bibnamefont{Whisnant}}, \bibinfo {author}
  {\bibfnamefont{S.}~\bibnamefont{Pakvasa}},\ and\ \bibinfo {author}
  {\bibfnamefont{R.}~\bibnamefont{Phillips}},\ }%
  \bibfield{journal}{%
  \Doi{10.1103/PhysRevD.22.2718}{\bibinfo {journal} {Phys. Rev. D}}\ }%
  \textbf{\bibinfo {volume} {22}},\ \bibinfo {pages} {2718} (\bibinfo {year}
  {1980})%
  \bibAnnoteFile{NoStop}{Barger:1980tf}%
\bibitem{Freund:1999gy}%
  \BibitemOpen
  \bibfield{author}{%
  \bibinfo {author} {\bibfnamefont{M.}~\bibnamefont{Freund}}, \bibinfo {author}
  {\bibfnamefont{M.}~\bibnamefont{Lindner}}, \bibinfo {author}
  {\bibfnamefont{S.}~\bibnamefont{Petcov}},\ and\ \bibinfo {author}
  {\bibfnamefont{A.}~\bibnamefont{Romanino}},\ }%
  \bibfield{journal}{%
  \Doi{10.1016/S0550-3213(00)00179-6}{\bibinfo {journal} {Nucl.Phys.}}\ }%
  \textbf{\bibinfo {volume} {B578}},\ \bibinfo {pages} {27} (\bibinfo {year}
  {2000})%
  \bibAnnoteFile{NoStop}{Freund:1999gy}%
\bibitem{Geer:2007kn}%
  \BibitemOpen
  \bibfield{author}{%
  \bibinfo {author} {\bibfnamefont{S.}~\bibnamefont{Geer}}, \bibinfo {author}
  {\bibfnamefont{O.}~\bibnamefont{Mena}},\ and\ \bibinfo {author}
  {\bibfnamefont{S.}~\bibnamefont{Pascoli}},\ }%
  \bibfield{journal}{%
  \Doi{10.1103/PhysRevD.75.093001}{\bibinfo {journal} {Phys. Rev. D}}\ }%
  \textbf{\bibinfo {volume} {75}},\ \bibinfo {pages} {093001} (\bibinfo {year}
  {2007})%
  \bibAnnoteFile{NoStop}{Geer:2007kn}%
\bibitem{Geer:1997iz}%
  \BibitemOpen
  \bibfield{author}{%
  \bibinfo {author} {\bibfnamefont{S.}~\bibnamefont{Geer}},\ }%
  \bibfield{journal}{%
  \Doi{10.1103/PhysRevD.57.6989, 10.1103/PhysRevD.59.039903}{\bibinfo {journal}
  {Phys. Rev. D}}\ }%
  \textbf{\bibinfo {volume} {57}},\ \bibinfo {pages} {6989} (\bibinfo {year}
  {1998})%
  \bibAnnoteFile{NoStop}{Geer:1997iz}%
\bibitem{DeRujula:1998hd}%
  \BibitemOpen
  \bibfield{author}{%
  \bibinfo {author} {\bibfnamefont{A.}~\bibnamefont{De~R\'ujula}}, \bibinfo
  {author} {\bibfnamefont{M.~B.}\ \bibnamefont{Gavela}},\ and\ \bibinfo
  {author} {\bibfnamefont{P.}~\bibnamefont{Hern\'andez}},\ }%
  \bibfield{journal}{%
  \Doi{10.1016/S0550-3213(99)00070-X}{\bibinfo {journal} {Nucl. Phys. B}}\ }%
  \textbf{\bibinfo {volume} {547}},\ \bibinfo {pages} {21} (\bibinfo {year}
  {1999})%
  \bibAnnoteFile{NoStop}{DeRujula:1998hd}%
\bibitem{Bandyopadhyay:2007kx}%
  \BibitemOpen
  \bibfield{author}{%
  \bibinfo {author} {\bibfnamefont{A.}~\bibnamefont{Bandyopadhyay}}
  \emph{et~al.} (\bibinfo {collaboration} {ISS Physics Working Group}),\ }%
  \bibfield{journal}{%
  \Doi{10.1088/0034-4885/72/10/106201}{\bibinfo {journal} {Rept. Prog. Phys.}}\
  }%
  \textbf{\bibinfo {volume} {72}},\ \bibinfo {pages} {106201} (\bibinfo {year}
  {2009})%
  \bibAnnoteFile{NoStop}{Bandyopadhyay:2007kx}%
\bibitem{Choubey:2011zz}%
  \BibitemOpen
  \bibfield{author}{%
  \bibinfo {author} {\bibfnamefont{S.}~\bibnamefont{Choubey}} \emph{et~al.},\
  }%
  \enquote{\bibinfo {title} {{International Design Study for the Neutrino
  Factory, Interim Design Report}},}\  (\bibinfo {year} {2011}),\ \bibinfo
  {note} {{IDS-NF-020}}%
  \bibAnnoteFile{NoStop}{Choubey:2011zz}%
\bibitem{Cervera:2010rz}%
  \BibitemOpen
  \bibfield{author}{%
  \bibinfo {author} {\bibfnamefont{A.}~\bibnamefont{Cervera}}, \bibinfo
  {author} {\bibfnamefont{A.}~\bibnamefont{Laing}}, \bibinfo {author}
  {\bibfnamefont{J.}~\bibnamefont{Martin-Albo}},\ and\ \bibinfo {author}
  {\bibfnamefont{F.~J.~P.}\ \bibnamefont{Soler}},\ }%
  \bibfield{journal}{%
  \Doi{10.1016/j.nima.2010.09.049}{\bibinfo {journal} {Nucl. Instrum. Meth.
  A}}\ }%
  \textbf{\bibinfo {volume} {624}},\ \bibinfo {pages} {601} (\bibinfo {year}
  {2010})%
  \bibAnnoteFile{NoStop}{Cervera:2010rz}%
\bibitem{Laing:2010zz}%
  \BibitemOpen
  \bibfield{author}{%
  \bibinfo {author} {\bibfnamefont{A.}~\bibnamefont{Laing}},\ }%
  Ph.D. thesis,\ \bibinfo {school} {{University of Glasgow}} (\bibinfo {year}
  {2010})%
  \bibAnnoteFile{NoStop}{Laing:2010zz}%
\bibitem{Rubbia:1977zz}%
  \BibitemOpen
  \bibfield{author}{%
  \bibinfo {author} {\bibfnamefont{C.}~\bibnamefont{Rubbia}},\ }%
  \enquote{\bibinfo {title} {{The Liquid Argon Time Projection Chamber: A New
  Concept for Neutrino Detectors}},}\  (\bibinfo {year} {1977}),\ \bibinfo
  {note} {{CERN-EP-INT-77-08}}%
  \bibAnnoteFile{NoStop}{Rubbia:1977zz}%
\bibitem{Cennini:1994br}%
  \BibitemOpen
  \bibfield{author}{%
  \bibinfo {author} {\bibfnamefont{P.}~\bibnamefont{Cennini}} \emph{et~al.}
  (\bibinfo {collaboration} {ICARUS~Collaboration}),\ }%
  \enquote{\bibinfo {title} {{ICARUS-II: a second generation proton decay
  experiment and neutrino observatory at the Gran Sasso Laboratory. Vol. I \&
  II}},}\  (\bibinfo {year} {1994}),\ \bibinfo {note} {{LNGS-94/99}}%
  \bibAnnoteFile{NoStop}{Cennini:1994br}%
\bibitem{Rubbia:2004tz}%
  \BibitemOpen
  \bibfield{author}{%
  \bibinfo {author} {\bibfnamefont{A.}~\bibnamefont{Rubbia}}\ }%
  \Eprint{http://arxiv.org/abs/hep-ph/0402110}{arXiv:hep-ph/0402110}%
  \bibAnnoteFile{NoStop}{Rubbia:2004tz}%
\bibitem{Rubbia:2009md}%
  \BibitemOpen
  \bibfield{author}{%
  \bibinfo {author} {\bibfnamefont{A.}~\bibnamefont{Rubbia}},\ }%
  \bibfield{journal}{%
  \Doi{10.1088/1742-6596/171/1/012020}{\bibinfo {journal} {J. Phys. Conf.
  Ser.}}\ }%
  \textbf{\bibinfo {volume} {171}},\ \bibinfo {pages} {012020} (\bibinfo {year}
  {2009})%
  \bibAnnoteFile{NoStop}{Rubbia:2009md}%
\bibitem{Bross:2007ts}%
  \BibitemOpen
  \bibfield{author}{%
  \bibinfo {author} {\bibfnamefont{A.~D.}\ \bibnamefont{Bross}}, \bibinfo
  {author} {\bibfnamefont{M.}~\bibnamefont{Ellis}}, \bibinfo {author}
  {\bibfnamefont{S.}~\bibnamefont{Geer}}, \bibinfo {author}
  {\bibfnamefont{O.}~\bibnamefont{Mena}},\ and\ \bibinfo {author}
  {\bibfnamefont{S.}~\bibnamefont{Pascoli}},\ }%
  \bibfield{journal}{%
  \Doi{10.1103/PhysRevD.77.093012}{\bibinfo {journal} {Phys. Rev. D}}\ }%
  \textbf{\bibinfo {volume} {77}},\ \bibinfo {pages} {093012} (\bibinfo {year}
  {2008})%
  \bibAnnoteFile{NoStop}{Bross:2007ts}%
\bibitem{Huber:2003ak}%
  \BibitemOpen
  \bibfield{author}{%
  \bibinfo {author} {\bibfnamefont{P.}~\bibnamefont{Huber}}\ and\ \bibinfo
  {author} {\bibfnamefont{W.}~\bibnamefont{Winter}},\ }%
  \bibfield{journal}{%
  \Doi{10.1103/PhysRevD.68.037301}{\bibinfo {journal} {Phys. Rev. D}}\ }%
  \textbf{\bibinfo {volume} {68}},\ \bibinfo {pages} {037301} (\bibinfo {year}
  {2003})%
  \bibAnnoteFile{NoStop}{Huber:2003ak}%
\bibitem{Smirnov:2006sm}%
  \BibitemOpen
  \bibfield{author}{%
  \bibinfo {author} {\bibfnamefont{A.}~\bibnamefont{Smirnov}}\ }%
  \Eprint{http://arxiv.org/abs/hep-ph/0610198}{arXiv:hep-ph/0610198}%
  \bibAnnoteFile{NoStop}{Smirnov:2006sm}%
\bibitem{Huber:2006wb}%
  \BibitemOpen
  \bibfield{author}{%
  \bibinfo {author} {\bibfnamefont{P.}~\bibnamefont{Huber}}, \bibinfo {author}
  {\bibfnamefont{M.}~\bibnamefont{Lindner}}, \bibinfo {author}
  {\bibfnamefont{M.}~\bibnamefont{Rolinec}},\ and\ \bibinfo {author}
  {\bibfnamefont{W.}~\bibnamefont{Winter}},\ }%
  \bibfield{journal}{%
  \Doi{10.1103/PhysRevD.74.073003}{\bibinfo {journal} {Phys. Rev. D}}\ }%
  \textbf{\bibinfo {volume} {74}},\ \bibinfo {pages} {073003} (\bibinfo {year}
  {2006})%
  \bibAnnoteFile{NoStop}{Huber:2006wb}%
\bibitem{Agarwalla:2010hk}%
  \BibitemOpen
  \bibfield{author}{%
  \bibinfo {author} {\bibfnamefont{S.~K.}\ \bibnamefont{Agarwalla}}, \bibinfo
  {author} {\bibfnamefont{P.}~\bibnamefont{Huber}}, \bibinfo {author}
  {\bibfnamefont{J.}~\bibnamefont{Tang}},\ and\ \bibinfo {author}
  {\bibfnamefont{W.}~\bibnamefont{Winter}},\ }%
  \bibfield{journal}{%
  \Doi{10.1007/JHEP01(2011)120}{\bibinfo {journal} {JHEP}}\ }%
  \textbf{\bibinfo {volume} {1101}},\ \bibinfo {pages} {120} (\bibinfo {year}
  {2011})%
  \bibAnnoteFile{NoStop}{Agarwalla:2010hk}%
\bibitem{FernandezMartinez:2010zza}%
  \BibitemOpen
  \bibfield{author}{%
  \bibinfo {author} {\bibfnamefont{E.}~\bibnamefont{Fern\'andez~Mart\'inez}},
  \bibinfo {author} {\bibfnamefont{T.}~\bibnamefont{Li}}, \bibinfo {author}
  {\bibfnamefont{S.}~\bibnamefont{Pascoli}},\ and\ \bibinfo {author}
  {\bibfnamefont{O.}~\bibnamefont{Mena}},\ }%
  \bibfield{journal}{%
  \Doi{10.1103/PhysRevD.81.073010}{\bibinfo {journal} {Phys. Rev. D}}\ }%
  \textbf{\bibinfo {volume} {81}},\ \bibinfo {pages} {073010} (\bibinfo {year}
  {2010})%
  \bibAnnoteFile{NoStop}{FernandezMartinez:2010zza}%
\bibitem{Huber:2008yx}%
  \BibitemOpen
  \bibfield{author}{%
  \bibinfo {author} {\bibfnamefont{P.}~\bibnamefont{Huber}}\ and\ \bibinfo
  {author} {\bibfnamefont{T.}~\bibnamefont{Schwetz}},\ }%
  \bibfield{journal}{%
  \Doi{10.1016/j.physletb.2008.10.009}{\bibinfo {journal} {Phys. Lett. B}}\ }%
  \textbf{\bibinfo {volume} {669}},\ \bibinfo {pages} {294} (\bibinfo {year}
  {2008})%
  \bibAnnoteFile{NoStop}{Huber:2008yx}%
\bibitem{Tang:2009wp}%
  \BibitemOpen
  \bibfield{author}{%
  \bibinfo {author} {\bibfnamefont{J.}~\bibnamefont{Tang}}\ and\ \bibinfo
  {author} {\bibfnamefont{W.}~\bibnamefont{Winter}},\ }%
  \bibfield{journal}{%
  \Doi{10.1103/PhysRevD.81.033005}{\bibinfo {journal} {Phys.Rev.}}\ }%
  \textbf{\bibinfo {volume} {D81}},\ \bibinfo {pages} {033005} (\bibinfo {year}
  {2010}),\ \Eprint{http://arxiv.org/abs/0911.5052}{arXiv:0911.5052 [hep-ph]}%
  \bibAnnoteFile{NoStop}{Tang:2009wp}%
\bibitem{Dighe:2011pa}%
  \BibitemOpen
  \bibfield{author}{%
  \bibinfo {author} {\bibfnamefont{A.}~\bibnamefont{Dighe}}, \bibinfo {author}
  {\bibfnamefont{S.}~\bibnamefont{Goswami}},\ and\ \bibinfo {author}
  {\bibfnamefont{S.}~\bibnamefont{Ray}}\ }%
  \Eprint{http://arxiv.org/abs/1110.3289}{arXiv:1110.3289 [hep-ph]}%
  \bibAnnoteFile{NoStop}{Dighe:2011pa}%
\bibitem{Pontecorvo:1957aa}%
  \BibitemOpen
  \bibfield{author}{%
  \bibinfo {author} {\bibfnamefont{B.}~\bibnamefont{Pontecorvo}},\ }%
  \bibfield{journal}{%
  \bibinfo {journal} {Zh.\ Eksp.\ Teor.\ Fiz.}\ }%
  \textbf{\bibinfo {volume} {33}},\ \bibinfo {pages} {549} (\bibinfo {year}
  {1957})%
  \bibAnnoteFile{NoStop}{Pontecorvo:1957aa}%
\bibitem{Pontecorvo:1958aa}%
  \BibitemOpen
  \bibfield{author}{%
  \bibinfo {author} {\bibfnamefont{B.}~\bibnamefont{Pontecorvo}},\ }%
  \bibfield{journal}{%
  \bibinfo {journal} {Zh.\ Eksp.\ Teor.\ Fiz.}\ }%
  \textbf{\bibinfo {volume} {34}},\ \bibinfo {pages} {247} (\bibinfo {year}
  {1958})%
  \bibAnnoteFile{NoStop}{Pontecorvo:1958aa}%
\bibitem{Mns:1962aa}%
  \BibitemOpen
  \bibfield{author}{%
  \bibinfo {author} {\bibfnamefont{Z.}~\bibnamefont{Maki}}, \bibinfo {author}
  {\bibfnamefont{M.}~\bibnamefont{Nakagawa}},\ and\ \bibinfo {author}
  {\bibfnamefont{S.}~\bibnamefont{Sakata}},\ }%
  \bibfield{journal}{%
  \bibinfo {journal} {Prog.\ Theor.\ Phys.}\ }%
  \textbf{\bibinfo {volume} {28}},\ \bibinfo {pages} {870} (\bibinfo {year}
  {1962})%
  \bibAnnoteFile{NoStop}{Mns:1962aa}%
\bibitem{Ankenbrandt:2009zza}%
  \BibitemOpen
  \bibfield{author}{%
  \bibinfo {author} {\bibfnamefont{C.}~\bibnamefont{Ankenbrandt}}, \bibinfo
  {author} {\bibfnamefont{S.~A.}\ \bibnamefont{Bogacz}}, \bibinfo {author}
  {\bibfnamefont{A.}~\bibnamefont{Bross}}, \bibinfo {author}
  {\bibfnamefont{S.}~\bibnamefont{Geer}}, \bibinfo {author}
  {\bibfnamefont{C.}~\bibnamefont{Johnstone}}, \emph{et~al.},\ }%
  \bibfield{journal}{%
  \Doi{10.1103/PhysRevSTAB.12.070101}{\bibinfo {journal} {Phys. Rev. ST Accel.
  Beams}}\ }%
  \textbf{\bibinfo {volume} {12}},\ \bibinfo {pages} {070101} (\bibinfo {year}
  {2009})%
  \bibAnnoteFile{NoStop}{Ankenbrandt:2009zza}%
\bibitem{Li:2010aa}%
  \BibitemOpen
  \bibfield{author}{%
  \bibinfo {author} {\bibfnamefont{T.}~\bibnamefont{Li}},\ }%
  Ph.D. thesis,\ \bibinfo {school} {Durham University} (\bibinfo {year}
  {2010})%
  \bibAnnoteFile{NoStop}{Li:2010aa}%
\bibitem{ONLINE}%
  \BibitemOpen
  \bibinfo {note} {\url{http://www.ippp.dur.ac.uk/~ballett/LENF}}%
  \bibAnnoteFile{NoStop}{ONLINE}%
\bibitem{Indumathi:2009hg}%
  \BibitemOpen
  \bibfield{author}{%
  \bibinfo {author} {\bibfnamefont{D.}~\bibnamefont{Indumathi}}\ and\ \bibinfo
  {author} {\bibfnamefont{N.}~\bibnamefont{Sinha}},\ }%
  \bibfield{journal}{%
  \Doi{10.1103/PhysRevD.80.113012}{\bibinfo {journal} {Phys. Rev. D}}\ }%
  \textbf{\bibinfo {volume} {80}},\ \bibinfo {pages} {113012} (\bibinfo {year}
  {2009})%
  \bibAnnoteFile{NoStop}{Indumathi:2009hg}%
\bibitem{Dutta:2011mc}%
  \BibitemOpen
  \bibfield{author}{%
  \bibinfo {author} {\bibfnamefont{R.}~\bibnamefont{Dutta}}, \bibinfo {author}
  {\bibfnamefont{D.}~\bibnamefont{Indumathi}},\ and\ \bibinfo {author}
  {\bibfnamefont{N.}~\bibnamefont{Sinha}}\ }%
  \Eprint{http://arxiv.org/abs/1103.5578}{arXiv:1103.5578 [hep-ph]}%
  \bibAnnoteFile{NoStop}{Dutta:2011mc}%
\bibitem{Nakamura:2010zzi}%
  \BibitemOpen
  \bibfield{author}{%
  \bibinfo {author} {\bibfnamefont{K.}~\bibnamefont{Nakamura}} \emph{et~al.}
  (\bibinfo {collaboration} {Particle Data Group}),\ }%
  \bibfield{journal}{%
  \Doi{10.1088/0954-3899/37/7A/075021}{\bibinfo {journal} {J. Phys. G}}\ }%
  \textbf{\bibinfo {volume} {37}},\ \bibinfo {pages} {075021} (\bibinfo {year}
  {2010})%
  \bibAnnoteFile{NoStop}{Nakamura:2010zzi}%
\bibitem{Donini:2010xk}%
  \BibitemOpen
  \bibfield{author}{%
  \bibinfo {author} {\bibfnamefont{A.}~\bibnamefont{Donini}}, \bibinfo {author}
  {\bibfnamefont{J.~J.}\ \bibnamefont{G\'omez~Cadenas}},\ and\ \bibinfo
  {author} {\bibfnamefont{D.}~\bibnamefont{Meloni}},\ }%
  \bibfield{journal}{%
  \Doi{10.1007/JHEP02(2011)095}{\bibinfo {journal} {JHEP}}\ }%
  \textbf{\bibinfo {volume} {1102}},\ \bibinfo {pages} {095} (\bibinfo {year}
  {2011})%
  \bibAnnoteFile{NoStop}{Donini:2010xk}%
\bibitem{Huber:2004ka}%
  \BibitemOpen
  \bibfield{author}{%
  \bibinfo {author} {\bibfnamefont{P.}~\bibnamefont{Huber}}, \bibinfo {author}
  {\bibfnamefont{M.}~\bibnamefont{Lindner}},\ and\ \bibinfo {author}
  {\bibfnamefont{W.}~\bibnamefont{Winter}},\ }%
  \bibfield{journal}{%
  \Doi{10.1016/j.cpc.2005.01.003}{\bibinfo {journal} {Comput. Phys. Commun.}}\
  }%
  \textbf{\bibinfo {volume} {167}},\ \bibinfo {pages} {195} (\bibinfo {year}
  {2005})%
  \bibAnnoteFile{NoStop}{Huber:2004ka}%
\bibitem{Huber:2007ji}%
  \BibitemOpen
  \bibfield{author}{%
  \bibinfo {author} {\bibfnamefont{P.}~\bibnamefont{Huber}}, \bibinfo {author}
  {\bibfnamefont{J.}~\bibnamefont{Kopp}}, \bibinfo {author}
  {\bibfnamefont{M.}~\bibnamefont{Lindner}}, \bibinfo {author}
  {\bibfnamefont{M.}~\bibnamefont{Rolinec}},\ and\ \bibinfo {author}
  {\bibfnamefont{W.}~\bibnamefont{Winter}},\ }%
  \bibfield{journal}{%
  \Doi{10.1016/j.cpc.2007.05.004}{\bibinfo {journal} {Comput. Phys. Commun.}}\
  }%
  \textbf{\bibinfo {volume} {177}},\ \bibinfo {pages} {432} (\bibinfo {year}
  {2007})%
  \bibAnnoteFile{NoStop}{Huber:2007ji}%
\bibitem{Dziewonski:1981aa}%
  \BibitemOpen
  \bibfield{author}{%
  \bibinfo {author} {\bibfnamefont{A.~M.}\ \bibnamefont{Dziewonski}}\ and\
  \bibinfo {author} {\bibfnamefont{D.~L.}\ \bibnamefont{Anderson}},\ }%
  \bibfield{journal}{%
  \Doi{10.1016/0031-9201(81)90046-7}{\bibinfo {journal} {Phys. Earth Planet.
  Inter.}}\ }%
  \textbf{\bibinfo {volume} {25}},\ \bibinfo {pages} {297} (\bibinfo {year}
  {1981})%
  \bibAnnoteFile{NoStop}{Dziewonski:1981aa}%
\bibitem{Stacey:1977aa}%
  \BibitemOpen
  \bibfield{author}{%
  \bibinfo {author} {\bibfnamefont{F.~D.}\ \bibnamefont{Stacey}},\ }%
  \emph{\bibinfo {title} {{Physics of the Earth}}},\ \bibinfo {edition} {2nd}\
  ed.\ (\bibinfo {publisher} {Wiley},\ \bibinfo {year} {1977})%
  \bibAnnoteFile{NoStop}{Stacey:1977aa}%
\bibitem{Cervera:2000kp}%
  \BibitemOpen
  \bibfield{author}{%
  \bibinfo {author} {\bibfnamefont{A.}~\bibnamefont{Cervera}}, \bibinfo
  {author} {\bibfnamefont{A.}~\bibnamefont{Donini}}, \bibinfo {author}
  {\bibfnamefont{M.~B.}\ \bibnamefont{Gavela}}, \bibinfo {author}
  {\bibfnamefont{J.~J.~G.}\ \bibnamefont{Cadenas}}, \bibinfo {author}
  {\bibfnamefont{P.}~\bibnamefont{Hern\'andez}}, \bibinfo {author}
  {\bibfnamefont{O.}~\bibnamefont{Mena}},\ and\ \bibinfo {author}
  {\bibfnamefont{S.}~\bibnamefont{Rigolin}},\ }%
  \bibfield{journal}{%
  \Doi{10.1016/S0550-3213(00)00221-2}{\bibinfo {journal} {Nucl. Phys. B}}\ }%
  \textbf{\bibinfo {volume} {579}},\ \bibinfo {pages} {17} (\bibinfo {year}
  {2000})%
  \bibAnnoteFile{NoStop}{Cervera:2000kp}%
\bibitem{Cervera:2001zz}%
  \BibitemOpen
  \bibfield{author}{%
  \bibinfo {author} {\bibfnamefont{A.}~\bibnamefont{Cervera}}, \bibinfo
  {author} {\bibfnamefont{A.}~\bibnamefont{Donini}}, \bibinfo {author}
  {\bibfnamefont{M.~B.}\ \bibnamefont{Gavela}}, \bibinfo {author}
  {\bibfnamefont{J.~J.~G.}\ \bibnamefont{Cadenas}}, \bibinfo {author}
  {\bibfnamefont{P.}~\bibnamefont{Hern\'andez}}, \bibinfo {author}
  {\bibfnamefont{O.}~\bibnamefont{Mena}},\ and\ \bibinfo {author}
  {\bibfnamefont{S.}~\bibnamefont{Rigolin}},\ }%
  \bibfield{journal}{%
  \Doi{10.1016/S0550-3213(00)00606-4}{\bibinfo {journal} {Nucl. Phys. B}}\ }%
  \textbf{\bibinfo {volume} {593}},\ \bibinfo {pages} {731} (\bibinfo {year}
  {2001})%
  \bibAnnoteFile{NoStop}{Cervera:2001zz}%
\bibitem{Drakoulakos:2004gn}%
  \BibitemOpen
  \bibfield{author}{%
  \bibinfo {author} {\bibfnamefont{D.}~\bibnamefont{Drakoulakos}} \emph{et~al.}
  (\bibinfo {collaboration} {Minerva Collaboration})\ }%
  \Eprint{http://arxiv.org/abs/hep-ex/0405002}{arXiv:hep-ex/0405002}%
  \bibAnnoteFile{NoStop}{Drakoulakos:2004gn}%
\bibitem{NOVA2007}%
  \BibitemOpen
  \bibfield{author}{%
  \bibinfo {author} {\bibnamefont{{NO$\nu$A Collaboration}}},\ }%
  \emph{\bibinfo {title} {{Technical Design Report}}} (\bibinfo {year} {2007})%
  \bibAnnoteFile{NoStop}{NOVA2007}%
\bibitem{Barger:2007jq}%
  \BibitemOpen
  \bibfield{author}{%
  \bibinfo {author} {\bibfnamefont{V.}~\bibnamefont{Barger}}, \bibinfo {author}
  {\bibfnamefont{P.}~\bibnamefont{Huber}}, \bibinfo {author}
  {\bibfnamefont{D.}~\bibnamefont{Marfatia}},\ and\ \bibinfo {author}
  {\bibfnamefont{W.}~\bibnamefont{Winter}},\ }%
  \bibfield{journal}{%
  \Doi{10.1103/PhysRevD.76.053005}{\bibinfo {journal} {Phys.Rev.}}\ }%
  \textbf{\bibinfo {volume} {D76}},\ \bibinfo {pages} {053005} (\bibinfo {year}
  {2007}),\ \Eprint{http://arxiv.org/abs/hep-ph/0703029}{arXiv:hep-ph/0703029
  [hep-ph]}%
  \bibAnnoteFile{NoStop}{Barger:2007jq}%
\bibitem{Bartoszek:2004si}%
  \BibitemOpen
  \bibfield{author}{%
  \bibinfo {author} {\bibfnamefont{L.}~\bibnamefont{Bartoszek}} \emph{et~al.}\
  }%
  \Eprint{http://arxiv.org/abs/hep-ex/0408121}{arXiv:hep-ex/0408121}%
  \bibAnnoteFile{NoStop}{Bartoszek:2004si}%
\bibitem{Cline:2006st}%
  \BibitemOpen
  \bibfield{author}{%
  \bibinfo {author} {\bibfnamefont{D.~B.}\ \bibnamefont{Cline}}, \bibinfo
  {author} {\bibfnamefont{F.}~\bibnamefont{Raffaelli}},\ and\ \bibinfo {author}
  {\bibfnamefont{F.}~\bibnamefont{Sergiampietri}},\ }%
  \bibfield{journal}{%
  \Doi{10.1088/1748-0221/1/09/T09001}{\bibinfo {journal} {JINST}}\ }%
  \textbf{\bibinfo {volume} {1}},\ \bibinfo {pages} {T09001} (\bibinfo {year}
  {2006})%
  \bibAnnoteFile{NoStop}{Cline:2006st}%
\bibitem{Baibussinov:2007ea}%
  \BibitemOpen
  \bibfield{author}{%
  \bibinfo {author} {\bibfnamefont{B.}~\bibnamefont{Baibussinov}}, \bibinfo
  {author} {\bibfnamefont{M.}~\bibnamefont{Baldo~Ceolin}}, \bibinfo {author}
  {\bibfnamefont{G.}~\bibnamefont{Battistoni}}, \bibinfo {author}
  {\bibfnamefont{P.}~\bibnamefont{Benetti}}, \bibinfo {author}
  {\bibfnamefont{A.}~\bibnamefont{Borio}}, \emph{et~al.},\ }%
  \bibfield{journal}{%
  \Doi{10.1016/j.astropartphys.2008.01.001}{\bibinfo {journal} {Astropart.
  Phys.}}\ }%
  \textbf{\bibinfo {volume} {29}},\ \bibinfo {pages} {174} (\bibinfo {year}
  {2008})%
  \bibAnnoteFile{NoStop}{Baibussinov:2007ea}%
\bibitem{Lipari:1994pz}%
  \BibitemOpen
  \bibfield{author}{%
  \bibinfo {author} {\bibfnamefont{P.}~\bibnamefont{Lipari}}, \bibinfo {author}
  {\bibfnamefont{M.}~\bibnamefont{Lusignoli}},\ and\ \bibinfo {author}
  {\bibfnamefont{F.}~\bibnamefont{Sartogo}},\ }%
  \bibfield{journal}{%
  \Doi{10.1103/PhysRevLett.74.4384}{\bibinfo {journal} {Phys. Rev. Lett.}}\ }%
  \textbf{\bibinfo {volume} {74}},\ \bibinfo {pages} {4384} (\bibinfo {year}
  {1995})%
  \bibAnnoteFile{NoStop}{Lipari:1994pz}%
\bibitem{Lipari:2002at}%
  \BibitemOpen
  \bibfield{author}{%
  \bibinfo {author} {\bibfnamefont{P.}~\bibnamefont{Lipari}},\ }%
  \bibfield{journal}{%
  \bibinfo {journal} {Nucl. Phys. Proc. Suppl.}\ }%
  \textbf{\bibinfo {volume} {112}},\ \bibinfo {pages} {274} (\bibinfo {year}
  {2002})%
  \bibAnnoteFile{NoStop}{Lipari:2002at}%
\bibitem{Dighe:2010js}%
  \BibitemOpen
  \bibfield{author}{%
  \bibinfo {author} {\bibfnamefont{A.}~\bibnamefont{Dighe}}, \bibinfo {author}
  {\bibfnamefont{S.}~\bibnamefont{Goswami}},\ and\ \bibinfo {author}
  {\bibfnamefont{S.}~\bibnamefont{Ray}},\ }%
  \bibfield{journal}{%
  \Doi{10.1103/PhysRevLett.105.261802}{\bibinfo {journal} {Phys. Rev. Lett.}}\
  }%
  \textbf{\bibinfo {volume} {105}},\ \bibinfo {pages} {261802} (\bibinfo {year}
  {2010})%
  \bibAnnoteFile{NoStop}{Dighe:2010js}%
\bibitem{Raut:2009jj}%
  \BibitemOpen
  \bibfield{author}{%
  \bibinfo {author} {\bibfnamefont{S.~K.}\ \bibnamefont{Raut}}, \bibinfo
  {author} {\bibfnamefont{R.~S.}\ \bibnamefont{Singh}},\ and\ \bibinfo {author}
  {\bibfnamefont{S.}~\bibnamefont{Uma~Sankar}},\ }%
  \bibfield{journal}{%
  \Doi{10.1016/j.physletb.2010.12.029}{\bibinfo {journal} {Phys. Lett. B}}\ }%
  \textbf{\bibinfo {volume} {696}},\ \bibinfo {pages} {227} (\bibinfo {year}
  {2011})%
  \bibAnnoteFile{NoStop}{Raut:2009jj}%
\bibitem{Joglekar:2010iu}%
  \BibitemOpen
  \bibfield{author}{%
  \bibinfo {author} {\bibfnamefont{A.}~\bibnamefont{Joglekar}}, \bibinfo
  {author} {\bibfnamefont{S.}~\bibnamefont{Prakash}}, \bibinfo {author}
  {\bibfnamefont{S.~K.}\ \bibnamefont{Raut}},\ and\ \bibinfo {author}
  {\bibfnamefont{S.}~\bibnamefont{Uma~Sankar}},\ }%
  \bibfield{journal}{%
  \Doi{10.1142/S0217732311036486}{\bibinfo {journal} {Mod. Phys. Lett. A}}\ }%
  \textbf{\bibinfo {volume} {26}},\ \bibinfo {pages} {2051} (\bibinfo {year}
  {2011})%
  \bibAnnoteFile{NoStop}{Joglekar:2010iu}%
\end{thebibliography}%
\end{document}